\documentclass[onecolumn]{elsart3p}
\usepackage{amssymb}
\usepackage{amsfonts}
\usepackage{amsmath}
\usepackage{graphicx}
\usepackage{subfigure}
\usepackage{epsfig}

\input{tcilatex}

\begin{document}

\begin{frontmatter}%

\title{Analytical approximation schemes for solving exact
renormalization group equations in the local potential approximation}%

\author{C.\ Bervillier}\ead{claude.bervillier@lmpt.univ-tours.fr},\author{ B.\ Boisseau}\ead{bruno.boisseau@lmpt.univ-tours.fr},\author{ H.\ Giacomini}\ead{hector.giacomini@lmpt.univ-tours.fr}%

\address{Laboratoire de Math\'{e}matiques et Physique Th\'{e}orique,\\ UMR 6083 (CNRS),\\
F\'ed\'eration Denis Poisson,\\
Universit\'{e} Fran\c{c}ois Rabelais,\\
Parc de Grandmont, 37200 Tours, France}
%

\begin{abstract}
The relation between the Wilson-Polchinski and the Litim optimized ERGEs in
the local potential approximation is studied with high accuracy using two
different analytical approaches based on a field expansion: a recently
proposed genuine analytical approximation scheme to two-point boundary value
problems of ordinary differential equations, and a new one based on
approximating the solution by generalized hypergeometric functions. A
comparison with the numerical results obtained with the shooting method is
made. A similar accuracy is reached in each case. Both two methods appear to
be more efficient than the usual field expansions frequently used in the
current studies of ERGEs (in particular for the Wilson-Polchinski case in
the study of which they fail).%
\end{abstract}%


\end{frontmatter}%

\section{Introduction}

The non-decoupling of the relevant scales on a wide and continuous range of
magnitudes in many areas of physics has led to the invention (discovery) of
the renormalisation group (RG) \cite{206}. Whereas they have been discovered
in the framework of the perturbative (quantum field) theory, the RG
techniques tackle a nonperturbative physical phenomenon \cite{425}.
Nonperturbative approaches are difficult to implement and to control, and
during a long time one has essentially carried on perturbative RG techniques
(see, e.g., \cite{4948}). Nowadays, the huge growth of the computing
capacity has greatly modified this behaviour pattern and, already since the
beginning of the ninety's, one has considered \cite{4374} with a greater
acuteness the exact RG equations (ERGEs) originally introduced by Wilson 
\cite{Irvine}, Wegner and Houghton \cite{414} in the seventy's and slightly
reformulated by Polchinski \cite{354} in the eighty's (for some reviews on
the ERGEs see \cite{4595}).

Initially, the ERGEs are integro-differential equations for the running
action $S\left[ \phi ,t\right] $ [assuming that $\phi \left( x\right) $
generically stands for some field with as many indices as necessary and $%
t=-\ln \left( \Lambda /\Lambda _{0}\right) $ the logarithm of a running
momentum scale $\Lambda $]. They have been extended to the running (average)
effective action $\Gamma \left[ \varphi ,t\right] $ \cite{4281,4374}. Such
general equations cannot be studied without the recourse to approximations
or truncations. One of the most promising approximations is a systematic
expansion in powers of the derivative of the field (derivative expansion) 
\cite{212} which yields a set of coupled nonlinear partial differential
equations the number of which grows quickly with the order of the expansion.
In the simplest cases (e.g., for the scalar field), the determination of
fixed points\ (and of their stability) amounts to study ordinary
differential equations (ODEs) with a two-point boundary value problem that
may be carried out numerically via a shooting (or a relaxation) method.

A pure numerical study is in general not easy to implement and to control.
For example, in the shooting method, the discovery of the right adjustment
of the parameters at the boundaries requires a good knowledge a priori of
their orders of magnitude (initial guesses). It is thus interesting to
develop concurrently some substitute analytical methods. A popular
substitute to the ODEs of the derivative expansion is provided by an
additionnal expansion in powers of the field which yields a set of coupled
algebraic equations which may be solved analytically, at least with the help
of a symbolic computation software. Various field expansions have been
implemented with more or less success \cite{3478,3642,4192,3553}.
Unfortunately, the methods proposed up to now, if they are easy to
implement, do not work in all cases and especially in the most famous and
simplest case of the Wilson-Polchinski ERGE \cite{Irvine,354} (equation for
the running action $S\left[ \phi ,t\right] $ with a smooth cutoff).

The object of this paper is to present two new substitute analytical methods
for studying ODEs which, at least in\ the local potential approximation of
the derivative expansion (LPA), works for the Wilson-Polchinski ERGE. One of
the methods, recently proposed in \cite{6110}, is a genuine analytical
approximation scheme to two-point boundary value problems of ODEs. The other
method is new. It is based on approximations of the solution looked for by
generalized hypergeometric functions. It has a certain similarity with
another new and interesting method based on the representation of the
solution by Pad\'{e} approximants just proposed in \cite{6201} by P. Amore
and F. M. Fernandez independantly from the present work. We illustrate the
effectiveness of the two methods with the explicit consideration of two
ERGEs in the local potential approximation: the Wilson-Polchinski equation
and the Litim optimized RG equation \cite{5020} for the running effective
action (named the Litim equation in the following).\ Following a conjecture
first stated in \cite{5049,5902}, the equivalence of these two equations (in
the LPA) has been proven by Morris \cite{5911} and recently been numerically
illustrated \cite{6137} with an unprecedented accuracy for the scalar field
in three dimensions ($d=3$). This particular situation provides us with the
opportunity of testing efficiently the various methods of study at hand.

The following of the paper is divided in five sections. In section \ref%
{Calculations}, we briefly present the direct numerical integration of the
ODEs for the scalar model\ using the shooting method: determinations of the
fixed point and the critical exponents for both the Wilson-Polchinski and
Litim equations in the LPA (distinguishing between the even and odd
symmetries). A brief presentation of the currently used field expansion is
given in section \ref{WPE}. In section \ref{BFG}, we analyse several aspects
of the method of \cite{6110} applying it to the study of the two equations.
We calculate this way the fixed point locations with high precision and
compare the results with the estimates obtained in section \ref{Calculations}%
. We show how the leading and the subleading critical exponents may be
estimated using this recent method. In section \ref{Ratios} we present a new
approximate analytical method for ODEs which is based on the definition of
the generalized hypergeometric functions. We show that it is well adapted to
treat the Wilson-Polchinski case whereas the Litim case is less easily
treated. We relate these effects to the convergence properties of the series
in powers of the field. Finally we summarize this work and conclude in
section \ref{Conc}.

\section{Two-point boundary value problem in the LPA}

\label{Calculations}

In this section we briefly present the two-point boundary value problem to
be solved in the LPA of the ERGE. The Wilson-Polchinski equation is first
chosen as a paradigm in section \ref{WilPol}. The principal numerical
results obtained from the numerical integration of the ODE using the
shooting method are given. In section (\ref{Litim}), the Litim equation is
also studied.

\subsection{Wilson-Polchinski's flow equation for the scalar-field\label%
{WilPol}}

The original Wilson-Polchinski ERGE in the LPA expresses the evolution of
the potential $U\left( \phi ,t\right) $ as varying the logarithm of the
momentum scale of reference $t=-\ln \left( \Lambda /\Lambda _{0}\right) $
(with $\phi \in 
\mathbb{R}
$). In three dimensions, it reads:%
\begin{equation}
\dot{U}=U^{\prime \prime }-\left( U^{\prime }\right) ^{2}-\frac{1}{2}\phi
U^{\prime }+3U\,,  \label{eq:LPAV}
\end{equation}%
in which $\dot{U}\equiv \partial U\left( \phi ,t\right) /\partial t$, $%
U^{\prime }\equiv \partial U\left( \phi ,t\right) /\partial \phi $, $%
U^{\prime \prime }\equiv \partial ^{2}U\left( \phi ,t\right) /\partial \phi
^{2}$.

\subsubsection{Fixed point equation\label{FP1}}

The fixed point equation corresponds to $\dot{U}=0$. It is a second order
ODE for the function $U\left( \phi \right) $:%
\begin{equation}
U^{\prime \prime }-\left( U^{\prime }\right) ^{2}-\frac{1}{2}\phi U^{\prime
}+3U=0\,,  \label{eq:FPF}
\end{equation}%
the solution of which (denoted $U^{\ast }\left( \phi \right) $ below)
depends on two integration constants which are fixed by two conditions. The
first one comes from a property of symmetry assumed to be\footnote{%
The other possibility $U^{\ast }\left( -\phi \right) =-U^{\ast }\left( \phi
\right) $ gives only singular solutions at finite $\phi $.} $U^{\ast }\left(
-\phi \right) =U^{\ast }\left( \phi \right) $ which provides the following
condition at the origin for $U^{\ast }\left( \phi \right) $:%
\begin{equation}
U^{\ast \prime }\left( 0\right) =0\,.  \label{eq:fori}
\end{equation}%
The second condition is the requirement that the solution we are interested
in must be non singular in the entire range $\phi \in \left[ 0,\infty \right[
$. Actually, the general solution of (\ref{eq:FPF}) involves a moving
singularity \cite{2080} of the form:%
\begin{equation}
U_{\text{sing}}=-\ln \left\vert \phi _{0}-\phi \right\vert \,,
\label{eq:sing}
\end{equation}%
depending on the arbitrary constant $\phi _{0}$. Pushing $\phi _{0}$ to
infinity allows to get a non-singular potential since, in addition to the
two trivial fixed points $U^{\ast }\equiv 0$ (Gaussian fixed point) and $%
U^{\ast }\equiv -\frac{1}{3}+\frac{{\phi }^{2}}{2}$ (high temperature fixed
point), eq.(\ref{eq:FPF}) admits a non-singular solution which, for $\phi
\rightarrow \infty $, has the form:%
\begin{equation}
U_{\text{asy}}(\phi )=\frac{{\phi }^{2}}{2}+b\,{\phi }^{\frac{6}{5}}+\frac{%
18\,b^{2}\,{\phi }^{\frac{2}{5}}}{25}-\frac{1}{3}+\frac{108\,b^{3}}{625\,{%
\phi }^{\frac{2}{5}}}+O\left( \phi ^{-4/5}\right) \,,  \label{eq:fasy}
\end{equation}%
in which $b$ is the only remaining arbitrary integration constant. The non
trivial (Wilson-Fisher \cite{439}) fixed point solution which we are
interested in must interpolate between eqs. (\ref{eq:fori}) and (\ref%
{eq:fasy}). Imposing these conditions fixes uniquely the value $b^{\ast }$
of $b$ which corresponds to the fixed point solution we are looking for$.$

We have determined $b^{\ast }$ by using the shooting method \cite{5465}:
starting from a value $\phi _{a}$ supposed to be large where the condition (%
\ref{eq:fasy}) is imposed (with a guess, or trying, value of $b\simeq
b^{\ast }$), we integrate the differential equation (\ref{eq:FPF}) toward
the origin where the condition (\ref{eq:fori}) is checked (shooting to the
origin), we adjust the value of $b$ to $b^{\ast }$ so as the latter
condition\ is satisfied with a required accuracy. A study of the stability
of the estimate of $b^{\ast }$ so obtained on varying the value $\phi _{a}$
provides some information on the accuracy of the calculation.

Rather than (\ref{eq:fasy}), it is more usual to characterize the fixed
point solution from its small field behaviour:

\begin{equation}
U(\phi )=k-\frac{3\,k\,}{2}\phi ^{2}+\frac{k\,\left( 1+3\,k\right) \,}{4}%
\phi ^{4}-\frac{k\,\left( 1+3\,k\right) \,\left( 1+24\,k\right) \,}{120}\phi
^{6}+O\left( \phi ^{8}\right) \,,  \label{eq:Uori}
\end{equation}%
and to provide the value of either of the two (related) quantities: 
\begin{eqnarray}
k^{\ast } &=&U^{\ast }\left( 0\right) \,,  \label{eq:kstarWP} \\
r^{\ast } &=&U^{\ast \prime \prime }\left( 0\right) =-3k^{\ast }\,.
\label{eq:rstar}
\end{eqnarray}

In the shooting-to-origin method, the determination of $r^{\ast }$ (or $%
k^{\ast }$)\ is a byproduct of the adjustment of $b^{\ast }$.

The adjustment of $b^{\ast }$ may be bypassed by shooting \emph{from} the
origin toward $\phi _{a}$, then $r^{\ast }$ is adjusted in such a way as to
reach the largest possible value of $\phi _{a}$. In that case $b^{\ast }$ is
a byproduct of the adjustment.

Because the boundary condition at $\phi _{a}$ is under control, the shooting-%
\emph{to}-origin method provides a better determination of $r^{\ast }$ than
the shooting-\emph{from}-origin method. However, this latter method is more
flexible and may easily yield a rough estimate on $r^{\ast }$ which can be
used as a guess in a more demanding management of the method. Notice that,
due to the increase of the number of adjustable parameters, this way of
determining a guess is no longer possible in a study involving several
coupled EDOs. Consequently, the development of other methods as, for
example, those two presented below is useful to this purpose (see also \cite%
{6201}).

\begin{table}[tbp]
\begin{center}
\begin{tabular}{ccccc}
\hline\hline
$r^{\ast }$ &  & $b^{\ast }$ &  & $\phi _{a}$ \\ \hline
\multicolumn{1}{l}{$-0.228\,598\,202\,437\,022\,0$} &  & \multicolumn{1}{l}{$%
\allowbreak -2.\,\allowbreak 296\,3$} &  & $10$ \\ 
\multicolumn{1}{l}{$-0.228\,598\,202\,437\,021\,9$} &  & \multicolumn{1}{l}{$%
\allowbreak -2.\,\allowbreak 311\,6$} &  & $20$ \\ 
\multicolumn{1}{l}{$-0.228\,598\,202\,437\,021\,9$} &  & \multicolumn{1}{l}{$%
\allowbreak -2.\,\allowbreak 316\,2$} &  & $40$ \\ \hline
\end{tabular}%
\end{center}
\caption{ The fixed point parameter $r^{\ast }$ is already well determined
for rather small values of $\protect\phi _{a}$ whereas $b^{\ast }$ [fixed
point value of $b$ in (\protect\ref{eq:fasy})] still is not. }
\label{Table 1}
\end{table}

Table \ref{Table 1} displays the determinations of $r^{\ast }$ and $b^{\ast
} $ for three values of $\phi _{a}$. One may observe that a high accuracy on 
$r^{\ast }$ is required to reach a yet small value of $\phi _{a}$ whereas $%
b^{\ast }$ is only poorly determined. Obviously, considering higher values
of $\phi _{a}$ and/or higher order terms in eq. (\ref{eq:fasy}) allows to
better determine $b^{\ast }$, one more term in (\ref{eq:fasy}) and $\phi
_{a}=1000$ yields:%
\begin{equation}
b^{\ast }=-2.318\,29\,,  \label{eq:bstar}
\end{equation}%
but the estimate of $r^{\ast }$ is not improved compared to the values given
in table \ref{Table 1} (the machine-precision was already reached). We
finally extract from table \ref{Table 1} our best estimate of $r^{\ast }$
(or $k^{\ast }$) as obtained from the study of the fixed point equation (\ref%
{eq:FPF}) alone:%
\begin{eqnarray}
r^{\ast } &=&-0.228\,598\,202\,437\,022\pm 10^{-15}\,\,,  \label{eq:rstar1}
\\
k^{\ast } &=&0.076\,199\,400\,812\,340\,7\pm 10^{-16}\,.  \label{eq:kstar1}
\end{eqnarray}

Individually, these values do not define the potential function $U^{\ast
}\left( \phi \right) $ the knowledge of which requires the numerical
integration explicitly performed in the shooting method.

\subsubsection{Eigenvalue equation}

The critical exponents are obtained by linearizing the flow equation (\ref%
{eq:LPAV}) near the fixed point solution $U^{\ast }\left( \phi \right) $. If
one inserts:%
\begin{equation*}
U\left( \phi ,t\right) =U^{\ast }\left( \phi \right) +\epsilon \,e^{\lambda
t}g\left( \phi \right) \,,
\end{equation*}%
into the flow equation and keeps the linear terms in $\epsilon $, one
obtains the eigenvalue equation:%
\begin{equation}
g^{\prime \prime }-2\,g^{\prime }U^{\ast \prime }-\frac{\phi }{2}\,g^{\prime
}+\left( 3-\lambda \right) \,g=0\,.  \label{eq:VPeq}
\end{equation}

Again it is a second order ODE the solutions of which are characterized by
two integration constants.

Since $U^{\ast }\left( \phi \right) $ is an even function of $\phi $, eq. (%
\ref{eq:VPeq}) is invariant under a parity change. Then one of the
integration constants is fixed by looking for either an even or an odd
eigenfunction $g\left( \phi \right) $ which implies either $g^{\prime
}\left( 0\right) =0$ (even) or $g\left( 0\right) =0$ (odd). The second
integration constant is fixed at will due to the arbitrariness of the
normalisation of an eigenfunction. Thus, assuming either $g\left( 0\right)
=1 $ (even) or $g^{\prime }\left( 0\right) =1$ (odd), the solutions of (\ref%
{eq:VPeq}) depend only on $\lambda $ and on the fixed point parameter $%
k^{\ast }$. For example, these solutions have the following expansions about
the origin $\phi =0$ :

\begin{eqnarray*}
&&g_{\text{even}}\left( \phi \right) =1+\frac{\left( \lambda -3\right) }{2}%
\phi ^{2}\left[ 1+\frac{\,\,\left( \lambda -2-12\,k^{\ast }\right) }{12}\phi
^{2}\right] +O\left( \phi ^{6}\right) \,, \\
&&g_{\text{odd}}\left( \phi \right) =\phi +\frac{\,\left( 2\,\lambda
-5-12\,k^{\ast }\right) }{12}\phi ^{3}+O\left( \phi ^{5}\right) \,.
\end{eqnarray*}%
When the fixed point solution $U^{\ast }$ is known, the values of $\lambda $
[the only remaining unknown parameter in (\ref{eq:VPeq})] are determined by
looking for the solutions which interpolate between either $g^{\prime
}\left( 0\right) =0$ (even) or $g\left( 0\right) =0$ (odd) and the regular
solution of (\ref{eq:VPeq}) \bigskip which, for $\phi \rightarrow \infty $,
is:%
\begin{equation}
g_{\text{asy}}(\phi )=S_{0}{\phi }^{\frac{2\,\left( 3-\lambda \right) }{5}%
}\left\{ 1+\left( 3-\lambda \right) \left[ \frac{12\,b^{\ast }\,}{25\,{\phi }%
^{\frac{4}{5}}}-\frac{36\,b^{\ast 2}\,\left( 2\,\lambda -3\right) }{625\,{%
\phi }^{\frac{8}{5}}}+\frac{2\,\,\left( 2\,\lambda -1\right) }{125\,{\phi }%
^{2}}+O\left( {\phi }^{-\frac{12}{5}}\right) \right] \right\} \,,
\label{eq:gasy}
\end{equation}%
in which $b^{\ast }$ is given by (\ref{eq:bstar}). The value of $S_{0}$ is
related to the choice of the normalisation of the eigenfunction at the
origin, it is a byproduct of the adjustment in a shooting-\emph{from}-origin
procedure.

In the even case, it is known that the first nontrivial positive eigenvalue $%
\lambda _{1}$ (there is also the trivial value $\lambda _{0}=d=3$), is
related to the critical exponent $\nu $ which characterizes the Ising-like
critical scaling of the correlation length $\xi $. One has $\nu =1/\lambda
_{1}$ and the first negative eigenvalue, $\lambda _{2}$, is minus the
Ising-like first correction-to-scaling exponent $\omega _{1}$ ($\omega
_{1}=-\lambda _{2}$) and so on.

In the odd case, the two first (positive) eigenvalues are trivial in the
LPA. One has:%
\begin{eqnarray}
\breve{\lambda}_{1} &=&\frac{d+2-\eta }{2}\,,  \label{eq:lambdac1} \\
\breve{\lambda}_{2} &=&\frac{d-2+\eta }{2}\,,  \label{eq:lambdac2}
\end{eqnarray}%
in which $\eta $ is the critical exponent which governs the large distance
behaviour of the correlation functions\ right at the critical point, it
vanishes in the LPA. With the dimension $d=3$ and the approximation (LPA)
presently considered, (\ref{eq:lambdac1}) and (\ref{eq:lambdac2}) reduce to $%
\breve{\lambda}_{1}=2.5$ and $\breve{\lambda}_{2}=0.5$. Consequently the
first non-trivial eigenvalue is negative and defines the subcritical
exponent $\theta _{5}=\breve{\omega}_{1}=-\breve{\lambda}_{3}$ sometimes
considered to characterize the deviation of the critical behaviour of fluids
from the pure Ising-like critical behaviour.

\begin{table}[tbp]
\begin{center}
\begin{tabular}{ccccc}
\hline\hline
$\nu $ &  & $b^{\ast }$ &  & $\phi _{a}$ \\ \hline
\multicolumn{1}{l}{$0.649\,561\,773\,880\,11$} &  & \multicolumn{1}{l}{$%
-2.\,\allowbreak 318\,145$} &  & \multicolumn{1}{l}{$12$} \\ 
\multicolumn{1}{l}{$0.649\,561\,773\,880\,80$} &  & \multicolumn{1}{l}{$%
-2.318\,257$} &  & \multicolumn{1}{l}{$22$} \\ 
\multicolumn{1}{l}{$0.649\,561\,773\,880\,65$} &  & \multicolumn{1}{l}{$%
-2.318\,280$} &  & \multicolumn{1}{l}{$32$} \\ 
\multicolumn{1}{l}{$0.649\,561\,773\,880\,65$} &  & \multicolumn{1}{l}{$%
-2.318\,285$} &  & \multicolumn{1}{l}{$40$} \\ \hline
\end{tabular}%
\end{center}
\caption{ Values of the critical exponent $\protect\nu $ determined together
with $b^{\ast }$ (and thus $r^{\ast }$) whereas $\protect\phi _{a}$ is
varied. Compared to table \protect\ref{Table 1}, a better determination of $%
b^{\ast }$ is obtained [see the best value of $b^{\ast }$ \ given by eq. (%
\protect\ref{eq:bstar})].}
\label{Table 2}
\end{table}

To determine the eigenvalues we use again the shooting-to-origin method with
the two equations (\ref{eq:FPF}, \ref{eq:VPeq}). However, in addition to $%
\lambda $, we leave also $b^{\ast }$ adjustable instead of fixing it to the
value given in (\ref{eq:bstar}).

In the even case, the values we obtain\ for $\nu $ and $b^{\ast }$ are shown
in table \ref{Table 2} for four values of $\phi _{a}$. Comparing with the
values displayed in table \ref{Table 1} one observes a better convergence of 
$b^{\ast }$ to the best value (\ref{eq:bstar}) whereas $r^{\ast }$ remains
unchanged compared to (\ref{eq:rstar1}). As for the best estimate of $\nu $,
it is:%
\begin{equation}
\nu _{\mathrm{best}}=0.649\,561\,773\,880\pm 10^{-12}\,,  \label{eq:nubest}
\end{equation}%
that is to say:%
\begin{equation}
\lambda _{1\mathrm{best}}=1.539\,499\,459\,808\pm 10^{-12}\,.
\label{eq:l1best}
\end{equation}

We have proceeded similarly to determine the Ising-like subcritical exponent
values displayed in table \ref{Table 3}.

\begin{table}[tbp]
\begin{center}
\begin{tabular}{ccccccccccc}
\hline\hline
$\omega _{1}$ &  & $\omega _{2}$ &  & $\omega _{3}$ &  & $\omega _{4}$ &  & $%
\omega _{5}$ &  & $\omega _{6}$ \\ \hline
$0.655\,745\,939\,193$ &  & $3.180\,006\,512\,059$ &  & $5.912\,230\,612$ & 
& $8.796\,092\,825$ &  & $11.798\,087\,66$ &  & $14.896\,053\,176$ \\ \hline
\end{tabular}%
\end{center}
\caption{ Best estimates of the six first subcritical exponents for the
Ising-like scalar model (i.e. even case), all digits are significant.}
\label{Table 3}
\end{table}

In the odd case, we obtain: 
\begin{equation}
\breve{\omega}_{1}=1.886\,703\,838\,091\pm 10^{-12}\,.  \label{eq:omegac}
\end{equation}

Table \ref{Table 4} displays the values of the other subcritical exponents
of the same family as $\breve{\omega}$ but with a lower accuracy. Of course,
the values presently obtained are in agreement with the previous estimates 
\cite{3491,6137}.

\begin{table}[tbp]
\begin{center}
\begin{tabular}{ccccc}
\hline\hline
$\breve{\omega}_{2}$ &  & $\breve{\omega}_{3}$ &  & $\breve{\omega}_{4}$ \\ 
\hline
$4.524\,390\,734$ &  & $7.337\,650\,643$ &  & $10.283\,900\,73$ \\ \hline
\end{tabular}%
\end{center}
\caption{ Best estimates of the odd-case subcritical exponents other than $%
\breve{\protect\omega}_{1}$ for the scalar model.}
\label{Table 4}
\end{table}

\subsection{ Litim's flow equation for the scalar field \label{Litim}}

Following a conjecture first stated in \cite{5049,5902}, the equivalence in
the LPA between the Wilson-Polchinski flow (\ref{eq:LPAV}) and the Litim
optimized ERGE \cite{5020} for the running effective action $\Gamma \left[
\varphi ,t\right] $ has been proven by Morris \cite{5911}. The Litim flow
equation for the potential $V\left( \varphi ,t\right) $ reads in three
dimensions (compared to \cite{5911} an unimportant shift $V\rightarrow V-1/3$
is performed):%
\begin{equation}
\dot{V}=1-\frac{1}{1+V^{\prime \prime }}-\frac{\varphi }{2}V^{\prime }+3V\,.
\label{eq:LPAVLitim}
\end{equation}

It is related to (\ref{eq:LPAV}) via the following Legendre transformation:%
\begin{equation}
\left. 
\begin{array}{l}
\left[ \frac{1}{2}\phi ^{2}-U\left( \phi ,t\right) \right] +\left[ \frac{1}{2%
}\varphi ^{2}+V\left( \varphi ,t\right) \right] =\varphi \phi \\ 
\varphi =\phi -U^{\prime }\left( \phi ,t\right)%
\end{array}%
\right\} \,.  \label{eq:Legendre}
\end{equation}

The general solution of the fixed point equation ($\dot{V}=0$) involves the
following moving \textquotedblleft singularity\textquotedblright\ ($%
V^{\prime \prime }$ is singular)\ at the arbitrary point $\varphi _{0}$:%
\begin{equation}
V_{\text{sing}}\left( \varphi \right) =-\frac{1}{3}+\frac{4}{3\sqrt{\varphi
_{0}}}\left\vert \varphi _{0}-\varphi \right\vert ^{3/2}\,.  \label{eq:singL}
\end{equation}

\subsubsection{Fixed point solution}

The numerical study of the fixed point solution of (\ref{eq:LPAVLitim})
follows the lines described in the preceding sections. This may be done
independently, but due to (\ref{eq:Legendre}), one may already deduce from
the previous study the expected results. Similarly to (\ref{eq:fasy}), the
asymptotic behaviour of the non trivial fixed point potential is
characterized by the integration constant $b_{L}$ in the following
expression [deduced from (\ref{eq:LPAVLitim})]:%
\begin{equation}
V_{\mathrm{asy}}(\varphi )=b_{L}\,{\varphi }^{6}-\frac{1}{3}+\frac{1}{%
150\,b_{L}\,{\varphi }^{4}}-\frac{1}{6300\,b_{L}^{2}\,{\varphi }^{8}}%
+O\left( {\varphi }^{-12}\right) \,.  \label{eq:asyv}
\end{equation}

It is easy to show from (\ref{eq:fasy}) and (\ref{eq:Legendre}) that the
value $b_{L}^{\ast }$ we are looking for is related to $b^{\ast }$ as
follows:%
\begin{equation*}
b_{L}^{\ast }\,=-\frac{1}{6^{6}}\left( \frac{5}{b^{\ast }}\right) ^{5},
\end{equation*}%
then, from the previous result (\ref{eq:bstar}) we get:%
\begin{equation}
b_{L}^{\ast }\simeq 0.001\,000\,25\,.  \label{eq:bstarL}
\end{equation}

Similarly for the potential parameters 
\begin{eqnarray*}
k_{L}^{\ast } &=&V^{\ast }\left( 0\right) \,, \\
r_{L}^{\ast } &=&V^{\ast \prime \prime }\left( 0\right) \,,
\end{eqnarray*}%
which correspond to $b_{L}^{\ast }$, they are related to the
Wilson-Polchinski counterparts $k^{\ast }$ and $r^{\ast }$ as follows:%
\begin{eqnarray}
k_{L}^{\ast } &=&k^{\ast }\,,  \label{eq:kstarL0} \\
r_{L}^{\ast } &=&\frac{r^{\ast }}{1-r^{\ast }}\,.  \label{eq:rstarL0}
\end{eqnarray}%
This latter relation, using (\ref{eq:rstar1}), gives:%
\begin{equation}
r_{L}^{\ast }\simeq -0.186\,064\,249\,470\,314\pm 10^{-15}\,.
\label{eq:rstarL}
\end{equation}

As precedingly, those values do not provide the potential function $V^{\ast
}\left( \varphi \right) $ the knowledge of which requires an explicit
numerical integration.

\subsubsection{Eigenvalue equation}

A linearization of the flow equation (\ref{eq:LPAVLitim}) near the fixed
point solution $V^{\ast }\left( \varphi \right) $:%
\begin{equation*}
V\left( \varphi ,t\right) =V^{\ast }\left( \varphi \right) +\epsilon
\,e^{\lambda t}h\left( \varphi \right) \,,
\end{equation*}%
provides the Litim eigenvalue equation:%
\begin{equation}
\left( 3-\lambda \right) \,h-\frac{\varphi \,h^{\prime }}{2}+\frac{h^{\prime
\prime }}{{\left( 1+{V^{\ast }}^{\prime \prime }\right) }^{2}}=0\,.
\label{eq:VPeqLitim}
\end{equation}

Taking into account (\ref{eq:asyv}), one can show that (\ref{eq:VPeqLitim})
admits a regular solution which, for $\varphi \rightarrow \infty $, has the
form:%
\begin{equation}
h_{\mathrm{asy}}\left( \varphi \right) =S_{1}{\varphi }^{2\,\left( 3-\lambda
\right) }\left\{ 1-\left( \lambda -3\right) \,\left( 2\,\lambda -5\right) %
\left[ \frac{1}{2250\,b_{L}^{\ast 2}\,{\varphi }^{10}}-\frac{1}{%
47250\,b_{L}^{\ast 3}\,{\varphi }^{14}}+O\left( \varphi ^{-18}\right) \right]
\right\} \,,  \label{eq:hasy}
\end{equation}%
in which $b_{L}^{\ast }$ is given by (\ref{eq:bstarL}). In the following we
may set $S_{1}=1$ since the normalisation of the eigenfunction may be chosen
at will.

As precedingly, we must distinguish between the odd and even eigenfunction $%
h\left( \varphi \right) $. The shooting method gives the same values as in
the Wilson-Polchinski case (see \cite{5049,5252,5625,6137}) and we do not
present them again.

\section{Expansion in powers of the field}

\label{WPE}

In advanced studies of the derivative expansion \cite{5469} or other
efficient approximations of the ERGE \cite{5903} and in the consideration of
complex systems via the ERGEs \cite{5677}, a supplementary truncation in
powers of the field is currently used (see also \cite{4595}). With a scalar
field, this expansion transforms the partial differential flow equations
into ODEs whereas the fixed point or eigenvalue ODEs are transformed into
algebraic equations. Provided auxiliary conditions are chosen, the latter
equations are easy to solve analytically using a symbolic computation
software. Actually the auxiliary conditions currently chosen are extremely
simple: they consist in setting equal to zero the highest terms of the
expansion so as to get a balanced system of equations.

A first kind of expansion, about the zero field --referred to as the
expansion I in the following, has been proposed by Margaritis et al \cite%
{3478} and applied to the LPA of Wegner-Houghton's ERGE \cite{414} (the hard
cutoff version of the Wilson-Polchinski equation). A second kind of
expansion, relative to the (running) minimum of the potential (expansion
II), has been proposed by Tetradis and Wetterich \cite{3642} and more
particularly presented by Alford \cite{4192} using it, again, with the sharp
cutoff version of the ERGE.

It is known that, for the Wegner-Houghton equation in the LPA, expansion I
does not converge due to the presence of singularities in the complex plane
of the expansion variable \cite{3358}. Expansions I and II have been more
concretely studied and compared to each other by Aoki et al in \cite{3553}
who also propose a variant to II (expansion III) by letting the expansion
point adjustable. They showed, again on the LPA of\ the Wegner-Houghton
equation, that expansion II is much more efficient than expansion I although
it finally does not converge and expansion III is the most efficient one.
Expansions II and III work well also on the ERGE expressed on the running
effective action (effective average action, see the review by Berges et al
in \cite{4595}). The convergence of those expansions have also been studied
in \cite{5252} according to the regularisation scheme chosen and in
particular for the Litim equation (\ref{eq:LPAVLitim}). In this latter study
it is concluded that both expansions I and II seem to converge although II
converges faster than I.

A striking fact emerges from those studies, the Wilson-Polchinski equation
in the LPA, the simplest equation, is never studied using the field
expansion method. The reason is simple: none of the expansions currently
used works in that case.

Actually the strategy of these methods, which consists in arbitrarily
setting equal to zero one coefficient for the expansion I and two for the
expansions II and III, is probably too simple. With regards to this kind of
auxiliary conditions, the failure observed with the Wilson-Polchinski
equation is not surprising and, most certainly, there should be many other
circumstances where such simple auxiliary conditions would not solve
correctly the derivative expansion of an ERGE.

In the following sections we examine two alternative methods with more
sophisticated auxiliary conditions. We show that they yield the correct
solution for the Wilson-Polchinski and its Legendre transformed (Litim)
equations. Both methods are associated to expansion I (about the
zero-field). The first one has recently been proposed in \cite{6110} as a
method to treat the two point boundary value problem of ODEs. It relies upon
an efficient account for the large field behaviour of the solution looked
for. An attempt of accounting for this kind of behaviour within the field
expansion had already been done by Tetradis and Wetterich via their eq.
(7.11) of \cite{3642}. In the present work, a much more sophisticated
procedure is used. It relies upon the construction of an added auxiliary
differential equation (ADE). We refer to it in the following as the ADE
method. The second method is new. It relies upon the approximation of the
solution looked for by a generalized hypergeometric function. We refer to it
in the following as the\ hypergeometric function approximation (HFA) method.

\section{Auxiliary differential equation method}

\label{BFG}

Let us first illustrate the auxiliary differential equation (ADE) method on
the search for the non trivial fixed point in the LPA for both the
Wilson-Polchinski equation (\ref{eq:FPF}) and the Litim optimized equation (%
\ref{eq:LPAVLitim}). Since there are two boundaries (the origin and the
"point at" infinity), we distinguish between two strategies.

\begin{itemize}
\item An expansion about the origin in the equations (small field expansion)
and the account for the leading high field behaviour of the regular solution
which we are looking for. This determines the value of $r^{\ast }$ or $%
r_{L}^{\ast }$.

\item A change of variable $\phi \rightarrow 1/\phi $ or $\varphi
\rightarrow 1/\varphi $ which reverses the problem: an expansion about
infinity (new origin) in the equations (high field expansion) and the
account for the leading small field behaviour of the regular solution which
we are looking for. This determines the value of $b^{\ast }$ or $b_{L}^{\ast
}$.
\end{itemize}

\subsection{Wilson-Polchinski's fixed point}

\subsubsection{ Small field expansion and leading high field behaviour\label%
{WPsmallField}}

For practical and custom reasons\footnote{%
The change $x=\phi ^{2}$ is useful in practice to avoid some degeneracies
observed in \cite{6110} when forming the auxiliary differential equation.
Taking the derivative $f=U^{\prime }$ is only a question of habit.}, instead
of (\ref{eq:FPF}) we consider the equation satisfied by the function $%
w\left( x\right) $ related to the derivative of the potential $U^{\prime
}\left( \phi \right) $ as follows:

\begin{equation}
U^{\prime }\left( \phi \right) =\phi \,w\left( \phi ^{2}\right) \,,
\label{eq:chgt1}
\end{equation}%
so that, with $x=\phi ^{2}$, the fixed point equation (\ref{eq:FPF}) reads:

\begin{equation}
4\,x\,w^{\prime \prime }-2\,{w}^{2}-4\,x\,w\,w^{\prime }+\left( 6-\,x\right)
\,\,w^{\prime }+2\,w=0\,,  \label{eq:FPw}
\end{equation}%
in which a prime indicates a derivative with respect to $x$.

This second order ODE has a singular point at the origin and, by analyticity
requirement, the solution we are looking for depends on a single unknown
integration-constant (noted $r$ below).

Let us first introduce the expansion I of Margaritis et al \cite{3478}. The
function $w\left( x\right) $ is expanded up to order $M$ in powers of $x$:%
\begin{equation}
w_{M}\left( x\right) =r+\sum\limits_{n=1}^{M}a_{n}x^{n}\,,  \label{eq:wm}
\end{equation}%
and inserted into the fixed point equation (\ref{eq:FPw}).

Requiring that (\ref{eq:FPw}) be satisfied order by order in powers of $x$
provides an unbalanced system of $M$ algebraic equations with $M+1$ unknown
quantities $\left\{ r,a_{1},\cdots ,a_{M}\right\} $ [eq. (\ref{eq:FPw}) is
then satisfied up to order $M-1$ in powers of $x$]. With a view to balancing
the system, $a_{M}=0$ is simply set equal to zero and if the solution
involves a stable value $r_{M}^{\ast }$ as $M$ grows, then it constitutes
the estimate at order $M$ of the fixed point location corresponding to
expansion I. As already mentioned, in the case of the Wilson-Polchinski
equation (\ref{eq:FPw}) under study, the method fails: all the values
obtained for $r_{M}^{\ast }$ are positive whatever the value of $M$ whereas
the correct value should be negative as shown in section \ref{FP1}.

In the ADE method, the condition $a_{M}=0$ is not imposed. The previous
algebraic system is first solved in terms of the unknown parameter $r$ \ so
as to get the generic solution of (\ref{eq:FPw}) at order $M$ in powers of $%
x $:%
\begin{equation}
w_{M}\left( r;x\right) =r+\sum\limits_{n=1}^{M}a_{n}\left( r\right) x^{n}\,.
\label{eq:wmrx}
\end{equation}

In order to get a definite value for $r$, instead of arbitrarily imposing $%
a_{M}\left( r\right) =0$, an auxiliary condition is formed which explicitly
accounts for the behaviour at large $\phi $ given by (\ref{eq:fasy}). With $%
w\left( x\right) $, this behaviour corresponds to:%
\begin{eqnarray}
&&w_{\text{asy}}(x)\underset{x\rightarrow \infty }{=}1\,,  \label{eq:asyw} \\
&&w_{\text{asy}}^{\prime }(x)\underset{x\rightarrow \infty }{=}0\,.
\label{eq:asyw'}
\end{eqnarray}%
The auxiliary condition is obtained via the introduction of an auxiliary
differential equation:

\begin{itemize}
\item Consider a first order differential equation for $w\left( x\right) $
constructed as a polynomial of degree $s$ (eventually incomplete) in powers
of the pair $\left( w,w^{\prime }\right) $: 
\begin{equation}
G_{1}+G_{2}\,w+G_{3}\,w^{\prime }+G_{4}\,w^{2}+G_{5}\,w\,w^{\prime
}+G_{6}w^{\prime 2}+\cdots +G_{n}\,w^{s-q}\,w^{\prime q}=0\,,
\label{eq:auxil1}
\end{equation}%
in which, when the degree $s$ of the polynomial is saturated then $q=s$ and
the number $n$ of coefficients $G_{i}$ is equal to $\left( s+1\right) \left(
s+2\right) /2$, conversely when it is not\ then $0\leq q<s$ and $%
n=s(s+1)/2+q+1$.

\item The constant coefficients $G_{i}$ are then determined as functions of $%
r$ by imposing that the solution $w_{M}\left( r;x\right) $ of (\ref{eq:FPw})
previously determined for arbitrary $r$ at order $M$ in powers of $x$ be
also solution of (\ref{eq:auxil1}) (at the same order $M$). Due to an
arbitrary normalisation which allows to fix, for example $G_{1}=1$, a simple
counting shows that the identification implies $M=n-1$. The resulting set $%
\{G_{i}\left( r\right) ;i=2,\ldots ,n\}$ is formed of rational functions of
the unknown parameter $r$. Hence, a new differential equation for $w\left(
x\right) $ is obtained:%
\begin{equation}
1+G_{2}\left( r\right) \,w+G_{3}\left( r\right) \,w^{\prime }+\,G_{4}\left(
r\right) \,w^{2}+G_{5}\left( r\right) \,w\,w^{\prime }+G_{6}\left( r\right)
\,w^{\prime 2}+\cdots +G_{n}\left( r\right) \,w^{s-q}\,w^{\prime q}=0\,,
\label{eq:auxilw}
\end{equation}%
which is satisfied by construction at order $M$ in powers of $x$ by (\ref%
{eq:wmrx}) which is already solution at the same order of (\ref{eq:FPw}).

\item \bigskip The last step is then to impose that the new equation (\ref%
{eq:auxilw}) be also satisfied when $x\rightarrow \infty $. Taking into
account (\ref{eq:asyw}, \ref{eq:asyw'}) it comes the final auxiliary
condition:%
\begin{equation}
1+G_{2}\left( r\right) \,+\,G_{4}\left( r\right) +\cdots +G_{s\left(
s+1\right) /2+1}\left( r\right) \,=0\,.  \label{eq:auxilCond}
\end{equation}
\end{itemize}

Solving this auxiliary condition for $r$ amounts to determining the roots of
a polynomial in $r$. As the order $M$ grows some root values appear to be
stable. Those stable values are candidates for the fixed point solutions we
are looking for. In a way similar to \cite{6201}, the obtention of the
auxiliary condition may be obtained without determinating explicitly the
coefficient functions $G_{i}\left( r\right) $. For this, it is sufficient to
consider the matrix $\mathcal{F}$ of the homogeneous system of linear
equations for all the $G_{i}$'s formed with eq (\ref{eq:auxil1}) to which is
added its expression when $x\rightarrow \infty $. When the function $w\left(
x\right) $ is replaced by the expansion (\ref{eq:wmrx}) at the required
order the matrix $\mathcal{F}$ depends only on the coefficients $a_{n}\left(
r\right) $ of the Taylor expansion (\ref{eq:wmrx}) and the auxiliary
condition then finally reduces to: 
\begin{equation}
\det \mathcal{F}=0  \label{eq:detF}
\end{equation}

Before going further, it is worthwhile indicating that a variant of the
method which consists in remplacing $w^{\prime }$ by $xw^{\prime }$ in the
auxiliary differential equation (\ref{eq:auxil1}) has appeared more
efficient [e.g., see figure \ref{fig5}]).

\begin{figure}[tbp]
\begin{center}
\includegraphics*[width=10cm]{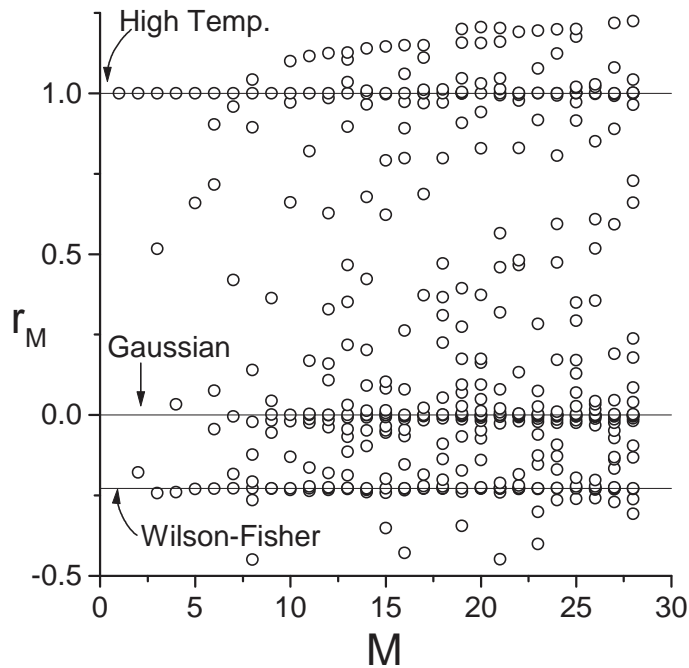}
\end{center}
\caption{Distribution of the real roots $r_{M}$ (open circles) of (\protect
\ref{eq:auxilCond}) as function of the order $M$ of the Taylor series about
the origin (\protect\ref{eq:wm}) [with the ADE pair $\left( w,xw^{\prime
}\right) $]. A threefold accumulation occurs about the expected fixed
points: trivial high temperature ($r^{\ast }=1$) and Gaussian ($r^{\ast }=0$%
)\ fixed points and about the non-trivial Wilson-Fisher fixed point ($%
r^{\ast }\simeq -0.2286$) [LPA, d=3].}
\label{fig2}
\end{figure}

\begin{figure}[tbp]
\begin{center}
\includegraphics*[width=10cm]{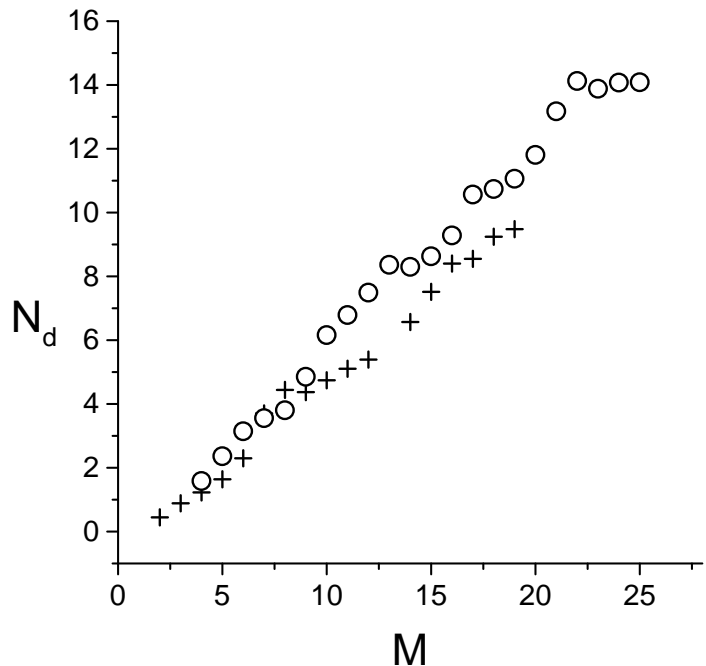}
\end{center}
\caption{Approximate number of accurate digits $N_{d}=-\log \left\vert
1-r_{M}/r^{\ast }\right\vert $ obtained on the selected roots $r_{M}$ as a
function of $M$ and for two ADE pairs; the original $\left( w,w^{\prime
}\right) $: crosses, and the variant $\left( w,xw^{\prime }\right) $: open
circles [$r^{\ast }$ is given in eq. (\protect\ref{eq:rstar1})]. The highest
values of $M$ in each case is limited by time computing. A better efficiency
is obtained with the variant.}
\label{fig5}
\end{figure}

Figure (\ref{fig2}) shows the distribution of all the real roots $r_{M}$ of (%
\ref{eq:auxilCond}) for the variant as the order $M$ varies up to 28. The
three expected fixed points encountered in section (\ref{FP1}) are clearly
evidenced by a threefold accumulation about the respective values $1$ (HT)$,$
$0$ (Gaussian) and $r^{\ast }$ (Wilson-Fisher). Although a huge accumulation
of roots around the right value occurs, the approach to $r^{\ast }$, which
we are interested in, may be followed step by step as the order $M$ grows. 

\paragraph{Selection of the root}

To select the right value of the root corresponding to the nontrivial
Wilson-Fisher fixed point, the following procedure has been applied. We know
that the root of interest is negative and real, then we select the first
negative real root that appears at the smallest possible order. At the next
order we choose the real root the closest to the previous choice and so on.
We obtain this way with $M=28$ the following excellent estimate: 
\begin{equation}
r^{\ast }=-0.228\,598\,202\,437\,02\,,  \label{eq:rstar2}
\end{equation}%
which coincides, up to the 14$^{th}$ digit, with the estimate (\ref%
{eq:rstar1}) obtained by the shooting method. Figure \ref{fig5} shows the
accuracy obtained on $r^{\ast }$ by selecting the roots this way as $M$
varies. 

\subsubsection{Subleading high field behaviour\label{SubWP}}

Equations (\ref{eq:asyw}, \ref{eq:asyw'}) used in the preceding calculations
express exclusively the limit of the solution $w^{\ast }\left( x\right) $
when $x\rightarrow \infty $, and we get the unique condition (\ref%
{eq:auxilCond}) to estimate $r^{\ast }$. In fact there are higher correction
terms to (\ref{eq:asyw}, \ref{eq:asyw'}) which vanish as $x\rightarrow
\infty $ [the first of which correspond to those written in (\ref{eq:fasy}%
)]. Such subleading contributions may as well be imposed in (\ref{eq:auxilw}%
). In so doing, we require the auxiliary differential equation to be
satisfied not only at infinity but also in approaching this point.
Consequently we obtain several auxiliary conditions similar to (\ref%
{eq:auxilCond}), each of them corresponding to the cancellation of the
coefficient of a given power of $x.$ We have used them to determine $r^{\ast
}$ again (the asymptotic constant $b$ factorizes in the first subleading
conditions so obtained). The results are similar to those obtained
precedently with the leading conditions (\ref{eq:asyw}, \ref{eq:asyw'})
alone. We have observed only a slight decrease in the accuracy: the higher
the subleading term considered the weaker the convergence to $r^{\ast }$.
This shows the coherence of the ADE method: the auxiliary condition is not
an isolated point condition, it emanates from a differential equation
constructed to be satisfied by the \emph{function} looked for.

When the order of the subleading contribution is high enough, the constant $%
b $ no longer factorizes and the subleading auxiliary condition depends non
trivialy on the (non-independent) integration constants ($r$ and $b$)
characterizing the fixed point solution. We have tried to determine the
value $b^{\ast }$ by imposing the individual vanishing of such contributions
for $r=r^{\ast }$. Unfortunately, at the orders considered, the only
knowledge of $r^{\ast }$ suffices to satisfy the condition (whatever the
value of $b)$. It is possible that considering much higher orders would
allow us to get an estimate of $b^{\ast }$ this way.

\subsubsection{ High field expansion and leading small field behaviour}

With the determination of $b^{\ast }$ by the ADE method in view, let us
perform the change of variable $x\rightarrow y^{-5}$ and the following
change of function:%
\begin{equation}
u\left( y\right) =y^{-2}\left[ w\left( y^{-5}\right) -1\right] \,,
\label{eq:chgt2}
\end{equation}%
so that, from (\ref{eq:fasy}) and (\ref{eq:chgt1}), $u^{\ast }\left(
y\right) $ has the following form for small $y:$%
\begin{equation}
u^{\ast }\left( y\right) =A^{\ast }+\frac{\,1}{5}A^{\ast 2}{y}^{2}-\frac{1}{%
25}A^{\ast 3}{y}^{4}+O\left( y^{5}\right) \,,  \label{eq:ustary}
\end{equation}%
with 
\begin{equation}
A^{\ast }=6b^{\ast }/5\,.  \label{eq:astar}
\end{equation}

The\ fixed point differential equation (\ref{eq:FPF}) is then transformed
into:

\begin{equation}
-10\,y\,{u}^{2}+5\,\left( 5+2\,y^{5}\right) \,u^{\prime }-4\,\left(
y^{4}-5\,y^{2}\,u^{\prime }\right) \,u+4\,y^{6}\,u^{\prime \prime }=0\,,
\label{eq:WFPasy}
\end{equation}%
the solution of which must satisfy the following condition, see (\ref%
{eq:ustary}):%
\begin{eqnarray*}
u^{\ast }\left( 0\right) &=&A^{\ast }\,, \\
u^{\ast \prime }\left( 0\right) &=&0\,,
\end{eqnarray*}%
with $A^{\ast }$ to be determined so as, using (\ref{eq:wm}, \ref{eq:chgt2}%
), to get at infinity:

\begin{equation*}
u_{\text{asy}}^{\ast }\left( y\right) \underset{y\rightarrow \infty }{=}0\,.
\end{equation*}

\begin{figure}[tbp]
\begin{center}
\includegraphics*[width=10cm]{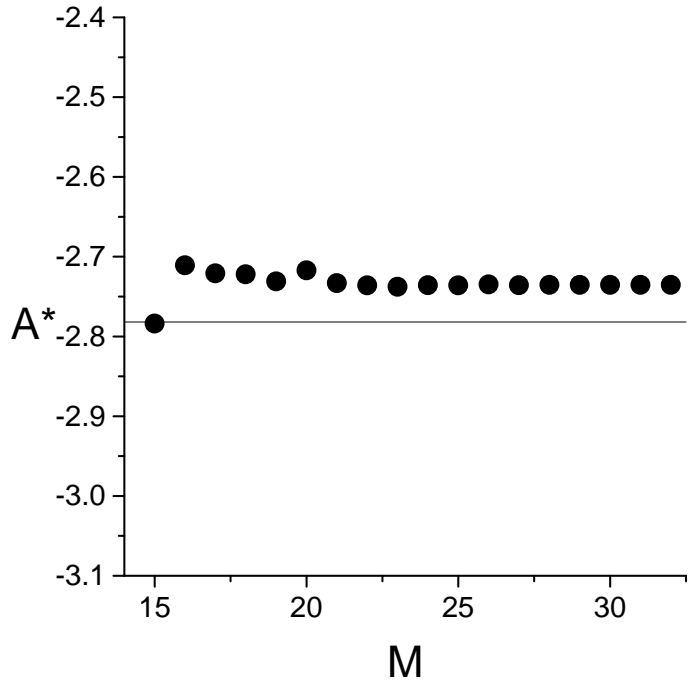}
\end{center}
\caption{ In the Wilson-Polchinski case, $A^{\ast }$ does not converge
(dots) to the right value $-2.78195$ (horizontal line) but to $-2.73532$.}
\label{fig3}
\end{figure}

The ADE method described in the preceding sections is used to determine the
value of $A^{\ast }$. Since there are some holes in the first terms of the
series (\ref{eq:ustary}), the first significant estimates are obtained for
values of $M$ higher than in section \ref{WPsmallField}. Figure (\ref{fig3})
shows that the selected sequence of roots corresponding to $A^{\ast }$
converges to $-2.73532$ whereas, according to (\ref{eq:bstar}, \ref{eq:astar}%
), the right value expected from the shooting method is $-2.78195$. 

This failure of the ADE method in determining correctly $A^{\ast }$ is
presumably due to the zero radius of convergence of the Taylor series of $%
u^{\ast }\left( y\right) $ about $y=0$. Actually, we have estimated this
radius as the limit of the ratio of two consecutive terms and observed that
it goes slowly but continuously to zero as the order $M$ increases. This
contrasts with the case of $w\left( x\right) $ for which the same procedure
quickly tends to the following finite limit for the fixed point solution
corresponding to (\ref{eq:rstar1}, \ref{eq:rstar2}):%
\begin{equation}
R_{WP}=5.721\,67\,.  \label{eq:RWP}
\end{equation}

Notice that, although the ADE method does not provide the right estimate of $%
A^{\ast }$ (or $b^{\ast }$), it gives a value close enough to it to be used
as a guess in the shooting method.

\subsection{Litim's fixed point}

\subsubsection{ Small field expansion and leading high field behaviour}

For convenience we perform the following change, compared to section \ref%
{Litim}:%
\begin{equation}
V\left( \varphi \right) =\bar{w}\left( \varphi ^{2}\right) -\frac{1}{3}\,,
\label{eq:Translation}
\end{equation}%
so that the fixed point equation corresponding to (\ref{eq:LPAVLitim}) reads
(with $\bar{x}=\varphi ^{2}$):

\begin{equation}
3\,\bar{w}-\bar{x}\,\bar{w}^{\prime }-\frac{1}{1+2\,\bar{w}^{\prime }+4\,%
\bar{x}\,\bar{w}^{\prime \prime }}=0\,.  \label{eq:wbarxbar}
\end{equation}

The singularity at $\bar{x}=0$ of this second order ODE allows us to look
for an analytic solution which satisfies, in terms of a single unknown
parameter $\bar{k}$, the following conditions at the origin:

\begin{eqnarray}
\bar{w}\left( 0\right) &=&\bar{k}\,,  \label{eq:wbarori1} \\
\bar{w}^{\prime }\left( 0\right) &=&\frac{1}{6\bar{k}}-\frac{1}{2}\,,
\label{eq:wbarori2}
\end{eqnarray}%
with $\bar{k}$ adjusted to $\bar{k}^{\ast }$ so as to reach at infinity
[from (\ref{eq:asyv})]:%
\begin{equation}
\bar{w}_{\mathrm{asy}}^{\ast }(\bar{x})=b_{L}^{\ast }\,{\bar{x}}^{3}+\frac{1%
}{150\,b_{L}^{\ast }\,{\bar{x}}^{2}}-\frac{1}{6300\,b_{L}^{\ast 2}\,{\bar{x}}%
^{4}}+O\left( {\bar{x}}^{-6}\right) \,.  \label{eq:asywbar}
\end{equation}

The expected value of $\bar{k}^{\ast }$ is related to $r_{L}^{\ast }$ given
in (\ref{eq:rstarL}) as:%
\begin{equation*}
\,\bar{k}^{\ast }=\frac{1}{3\left( 1+\,r_{L}^{\ast }\right) }\,.
\end{equation*}

It is also related to $k^{\ast }$ given in (\ref{eq:kstar1}) via (\ref%
{eq:kstarL0}, \ref{eq:Translation}) as $\bar{k}^{\ast }=k^{\ast }+1/3$.
Consequently the estimation by the shooting method is: 
\begin{equation}
\,\bar{k}^{\ast }=0.409\,532\,734\,\allowbreak 145\,674\pm 10^{-15}\,.
\label{eq:kstar}
\end{equation}

The object of this section is thus to test whether the ADE method yields
that value of $\,\bar{k}^{\ast }$ [and also that of $b_{L}^{\ast }$ given in
(\ref{eq:bstarL})].

Contrary to the Wilson-Polchinski case, the asymptotic behaviour (\ref%
{eq:asywbar}) does not reach a finite value when $\bar{x}\rightarrow \infty $%
. But the third derivative of $\bar{w}^{\ast }$ does. Hence, since $%
b_{L}^{\ast }$ is still supposed unknown, the auxiliary first order
differential equation (\ref{eq:auxil1}) may be used with $w$ and $w^{\prime
} $ replaced respectively by $\bar{w}^{(4)}$ and $\bar{w}^{(5)}$ (where $%
\bar{w}^{(n)}$ stands for d$^{n}\bar{w}/$d$\bar{x}^{n}$). Actually, both of
these\ two derivatives go to zero as $\bar{x}\rightarrow \infty $ so that
finally the auxiliary condition similar to (\ref{eq:auxilCond}), but with
another normalisation of the $G_{i}$'s (e.g. $G_{2}=1$), reduces to:%
\begin{equation}
G_{1}\left( \bar{k}\right) =0\,,  \label{eq:auxilCondL}
\end{equation}%
whereas the function $\bar{w}\left( \bar{x}\right) $ is expanded up to order 
$M$ in powers of $\bar{x}$ and inserted into (\ref{eq:wbarxbar}) to get the
solution at this order as function of $\bar{k}$:%
\begin{equation}
\bar{w}_{M}\left( \bar{k};\bar{x}\right) =\bar{k}+\sum\limits_{n=1}^{M}a_{n}%
\left( \bar{k}\right) \bar{x}^{n}\,.  \label{eq:wbarm}
\end{equation}

Similarly to the Wilson-Polchinski case, the complete set of real roots of (%
\ref{eq:auxilCondL}) shows accumulations about the expected fixed point
values. However the selection process described previously fails in picking
the right value $\bar{k}^{\ast }$ (of the nontrivial fixed point) although
it is present among the roots. Actually, for $M=14$ the selection gives $%
0.409\,627\,819\,729\,71$ whereas a better value ($0.409\,532\,733\,212\,35$%
) exists at the same order [compare with (\ref{eq:kstar})]. The variant
utilised in the preceding case which consists in replacing $\bar{w}^{(5)}$
by $\bar{x}\bar{w}^{(5)}$ does not circumvents this difficulty.

If instead of $\left( \bar{w}^{(4)},\bar{w}^{(5)}\right) $ as ADE pair, we
consider the combination $h=3\,\bar{w}-\bar{x}\,\bar{w}^{\prime }$ and its
derivative $h^{\prime }$ with respect to $\bar{x}$ (or the variant $\bar{x}%
h^{\prime }$ to save some time computing), then the new pair, according to (%
\ref{eq:asywbar}), vanishes also as $\bar{x}\rightarrow \infty $ , and we
observe, this time, that the selection process works again. This way, at
order $M=19$ the selection gives:%
\begin{equation*}
\bar{k}^{\ast }=0.409\,532\,734\,16\,,
\end{equation*}%
a value which coincides with (\ref{eq:kstar}) up to the 10$^{th}$ digit. No
doubt that considering higher values of $M$ would have improved the
accuracy. We note that, as with Wilson-Polchinski's function, the radius of
convergence of the Taylor series of $\bar{w}\left( \bar{x}\right) $ about
the origin is finite, and is about:%
\begin{equation}
R_{L}\simeq 11.5\,.  \label{eq:radiusL}
\end{equation}

Let us specify however that, contrary to the Wilson-Polchinski case, the
test of the ratio $a_{i}/a_{i+1}$ of two consecutive terms\ of the Taylor
series about the origin does not converge. We have obtained (\ref{eq:radiusL}%
) by explicitly performing a partial summation of the series and studying it
as a function of $\bar{x}$. Nevertheless, we have also observed that the
ratio $\left\vert a_{i}/a_{i+3}\right\vert $ raised to the power $1/3$,
roughly converges to (\ref{eq:radiusL}). This remark will have some
importance in section \ref{GeoLitim}.

\begin{figure}[tbp]
\begin{center}
\includegraphics*[width=10cm]{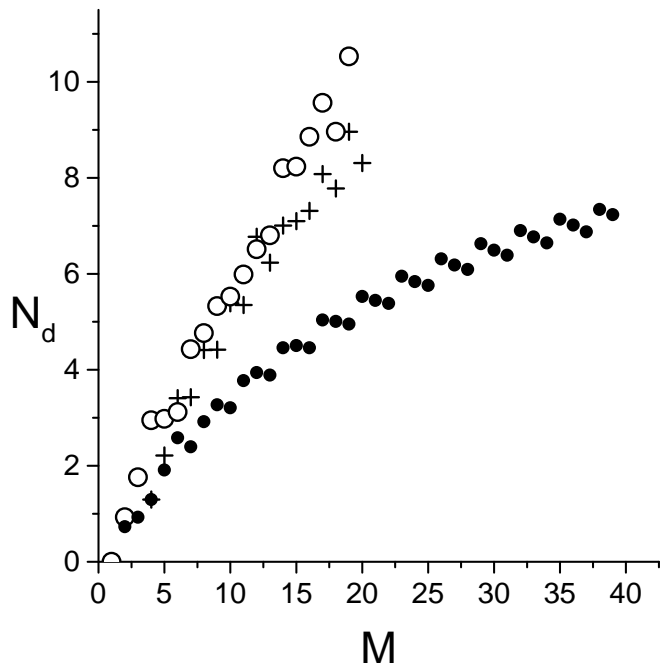}
\end{center}
\caption{Approximate number of accurate digits $N_{d}=-\log \left\vert 1-%
\bar{k}_{M}/\bar{k}^{\ast }\right\vert $ [with $\bar{k}^{\ast }$ given by (%
\protect\ref{eq:kstar})] as functions of $M$ for the estimations of the
Litim fixed point value of $\bar{k}$ using three methods: expansion I (black
dots), expansion II (crosses), and ADE (open circles). A better efficiency
is obtained with the ADE method.}
\label{fig1}
\end{figure}

Since expansions I and II work in the Litim case (see \cite{5252}), we can
compare the ADE method with those two methods. Figure (\ref{fig1}) shows the
respective accuracies obtained on $\bar{k}^{\ast }$ with the three methods
as functions of the order $M$ of the field expansion. One sees that
expansion II and the ADE method provide better results than expansion I
(which likely does not converge) and that the ADE method is most efficient
than expansion II (we have not studied expansion III). 

\subsubsection{Subleading high field behaviour}

As in the case of Wilson-Polchinski's equation, the subleading terms in (\ref%
{eq:asywbar}) may be used to impose the auxiliary condition not only at
infinity but also in approaching this point whatever the value of $\bar{x}$.
We observe the same phenomenon as in section \ref{SubWP}: the higher the
subleading term considered the weaker the convergence to $\bar{k}^{\ast }$
whereas $b_{L}^{\ast }$ cannot be determined by imposing the individual
vanishing of the subleading contributions for $\bar{k}=\bar{k}^{\ast }$.

However, the fact that the asymptotic behaviour (\ref{eq:asywbar}) is an
integer power of $\bar{x}$ provides us with the oportunity of determining $%
b_{L}^{\ast }$ from the knowledge of $\bar{k}^{\ast }$ as a boundary limit
(a point condition). Actually, since $\bar{w}^{(3)}\rightarrow 6b_{L}^{\ast
}\,$ when $\bar{x}\rightarrow \infty $, we may choose $\left( \bar{w}^{(3)},%
\bar{w}^{(4)}\right) $ as ADE pair [or the variant $\left( \bar{w}^{(3)},%
\bar{x}\bar{w}^{(4)}\right) $], and for $\bar{k}$ fixed to $\bar{k}^{\ast }$
solve for $b_{L}^{\ast }$ the resulting auxiliary condition at infinity. The
accuracy on $b_{L}^{\ast }$ obtained this way is not as large as in the case
of $\bar{k}^{\ast }$, nevertheless, for $M=31$ we obtain the following
estimation:%
\begin{equation}
b_{L}^{\ast }\simeq 0.001\,007\,,  \label{eq:bstarLsub}
\end{equation}%
which is rather close to the shooting value (\ref{eq:bstarL}). We indicate
also that rough estimates of $b_{L}^{\ast }$ already sufficiently accurate
to be used as guesses in the shooting method are obtained for small values
of $M$, e.g.: $0.000\,989$ for $M=11$ or even $0.0012$ for $M=5$.

\subsubsection{ High field expansion and leading small field behaviour}

With a view to determining $b_{L}^{\ast }$ directly by the ADE method, we
invert the boundaries by changing the variable $\bar{x}\rightarrow \bar{y}%
^{-1}$ and by performing the following change of function:%
\begin{equation}
\bar{u}\left( \bar{y}\right) ={\bar{y}}^{3}\bar{w}\left( \frac{1}{\bar{y}}%
\right) \,,  \label{eq:ubar}
\end{equation}%
so that, from (\ref{eq:asywbar}), we deduce that $\bar{u}^{\ast }\left( \bar{%
y}\right) $ has the following form for small $\bar{y}:$%
\begin{equation}
\bar{u}^{\ast }\left( \bar{y}\right) =b_{L}^{\ast }\,+\frac{{\bar{y}}^{5}}{%
150\,b_{L}^{\ast }\,}-\frac{{\bar{y}}^{7}}{6300\,b_{L}^{\ast 2}\,}+O\left( {%
\bar{y}}^{9}\right) \,.  \label{eq:ubarstarori}
\end{equation}

The differential equation for $\bar{u}\left( \bar{y}\right) $ is:

\begin{equation*}
\bar{y}^{4}+18\,\bar{y}\,\left( \bar{u}^{\prime }\right) ^{2}-\bar{u}%
^{\prime }\left[ 30\,\bar{u}+\bar{y}^{2}\,\left( 1+4\,\bar{u}^{\prime \prime
}\right) \right] =0\,.
\end{equation*}%
\bigskip

The solution must satisfy the following condition at the origin $\bar{y}=0$
[see (\ref{eq:ubarstarori})]:%
\begin{eqnarray*}
\bar{u}^{\ast }\left( 0\right) &=&b_{L}^{\ast }\,, \\
\bar{u}^{\ast \prime }\left( 0\right) &=&0\,,
\end{eqnarray*}%
with $b_{L}^{\ast }$ to be determined so as, using (\ref{eq:wbarori1}, \ref%
{eq:wbarori2}, \ref{eq:ubar}), to get at infinity:

\begin{equation*}
\bar{u}_{\text{asy}}^{\ast }\left( \bar{y}\right) =\bar{k}^{\ast }{\bar{y}}%
^{3}+\left( \frac{1}{6\bar{k}^{\ast }}-\frac{1}{2}\right) {\bar{y}}%
^{2}+O\left( \bar{y}\right) \,.
\end{equation*}

As previously, we use the ADE method with a view to determining the value of 
$b_{L}^{\ast }\,$. For this we consider, the pair $\left( \bar{u}^{(4)},\bar{%
u}^{(5)}\right) $ which vanishes at infinity ($\bar{y}\rightarrow \infty $).
Since there are some holes in the first terms of the series about the
origin, see (\ref{eq:ubarstarori}), the first significant estimates are
obtained for values of $M$ higher than with the original function $\bar{w}%
\left( \bar{x}\right) $. Although the positive roots obtained for $b_{L}$
(we know that $b_{L}^{\ast }$ is positive)\ have the right order of
magnitude compared to (\ref{eq:bstarL}) the apparent convergent sequences do
not provide the right value. Again, as in the Wilson-Polchinski case, we
think that the failure of the ADE method is due to the (observed) zero
radius of convergence of the Taylor series for $\bar{u}\left( \bar{y}\right) 
$ about the origin.

\subsection{Eigenvalue estimates}

Let us consider the eigenvalue problem with the ADE method. This time two
coupled nonlinear ODEs have to be solved together (the fixed point equation
and the linearisation of the flow in the vicinity of the fixed point). We
can solve these two equations together as the order of the field expansion $%
M $ grows or consider separately the eigenvalue equation after having solved
the fixed point equation with some accuracy. With the aim to be short, we
present only the latter possibility which illustrates well the property of
convergence of the method.

\subsubsection{Wilson-Polchinski's eigenvalues}

\paragraph{ Small field expansion and leading high field behaviour}

Using a change of eigenfunction, $g\rightarrow \mathrm{v}$, similar to (\ref%
{eq:chgt1}) for the fixed point function, it comes:

\begin{itemize}
\item in the even case:%
\begin{equation*}
g^{\prime }\left( \phi \right) =\phi \,\mathrm{v}\left( \phi ^{2}\right) \,,
\end{equation*}

and eq. (\ref{eq:gasy}) yields the following behaviour at large $x=\phi ^{2}$%
:%
\begin{equation*}
\mathrm{v}_{\text{asy}}\left( x\right) =\frac{2\left( 3-\lambda \right) }{5}%
S_{0}\,x^{-\left( 2+\lambda \right) /5}\left[ 1+O\left( x^{-2/5}\right) %
\right] \,.
\end{equation*}

\item in the odd case:%
\begin{equation*}
g^{\prime }\left( \phi \right) =\,\mathrm{v}\left( \phi ^{2}\right) \,,
\end{equation*}%
and eq. (\ref{eq:gasy}) gives:%
\begin{equation*}
\mathrm{v}_{\text{asy}}\left( x\right) =\frac{2\left( 3-\lambda \right) }{5}%
S_{0}\,x^{\left( 1-2\lambda \right) /10}\left[ 1+O\left( x^{-2/5}\right) %
\right] \,.
\end{equation*}
\end{itemize}

\begin{figure}[tbp]
\begin{center}
\includegraphics*[width=10cm]{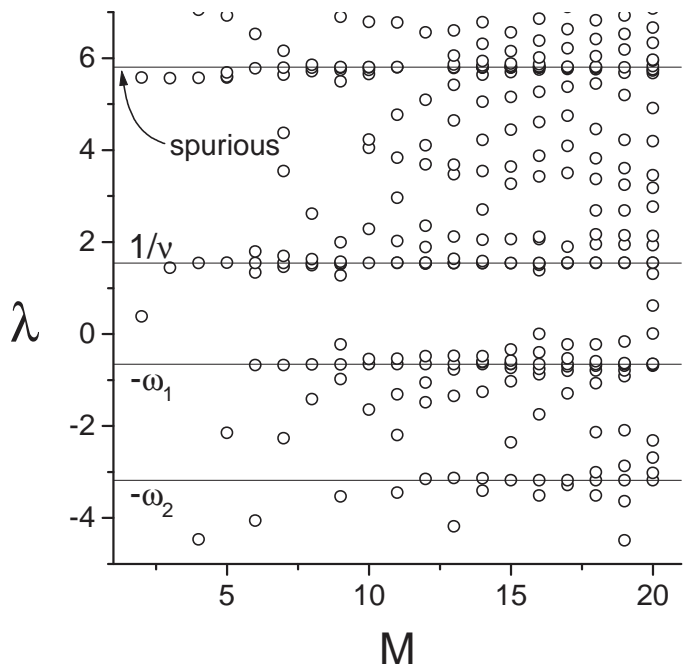}
\end{center}
\caption{Accumulations of real roots (open circles)\ of the auxiliary
condition (\protect\ref{eq:auxilCond}) about eigenvalues as the order $M$\
of the series varies in the Wilson-Polchinski even case. From top to bottom: 
$\protect\lambda _{1}=1/\protect\nu $ (second horizontal line), $\protect%
\lambda _{2}=-\protect\omega _{1}$ (third h. line) and $\protect\lambda %
_{3}=-\protect\omega _{2}$ (fourth h. line). A simple criterion of choice
allows to determine their estimates at $M=20$, see the values in eqs. (%
\protect\ref{eq:WPvpnu}--\protect\ref{eq:WPvpomega2}). An accumulation also
occurs about the spurious value 5.8 (first h. line).}
\label{fig4}
\end{figure}

The arbitrariness of the global normalisation of the eigenfunctions allows
to choose $\mathrm{v}\left( 0\right) =1$ (even) and $\mathrm{v}^{\prime
}\left( 0\right) =1$ (odd) corresponding respectively to some definite
values of $S_{0}$. So defined, the functions $\mathrm{v}\left( x\right) $
and $\mathrm{v}^{\prime }\left( x\right) $ vanish at infinity provided that $%
\lambda >-2$ in the even case and $\lambda >1/2$ in the odd case. Hence one
could expect that, with the simple condition at infinity: $\mathrm{v}=%
\mathrm{v}^{\prime }=0$ imposed in the auxiliary differential equation, the
ADE procedure will, at best, allow the determination of exclusively the
leading ($\lambda _{1}=1/\nu $) and first subleading ($\lambda _{2}=-\omega
_{1}$) eigenvalues in the even case and of only the trivial eigenvalue $%
\breve{\lambda}_{1}=-\breve{\omega}_{1}$ in the odd case [see the values of
these quantities in eqs. (\ref{eq:l1best}, \ref{eq:omegac}) and tables (\ref%
{Table 3}, \ref{Table 4})]. Actually it is better than that since, as $M$
grows, we observe\ among the real roots of the auxiliary condition for $%
\lambda $ that a hierarchy of successive accumulations takes place about the
right values of the leading and subsequent eigenvalues [see figure \ref{fig4}%
]. 

Within each of these accumulations of real roots, we have been able to
follow without ambiguity a convergent sequence to the right estimate. At
order $M=20$ with the ADE pair $\left( \mathrm{v},\mathrm{v}^{\prime
}\right) $ supposed to vanish at infinity, and $r^{\ast }$ fixed to the
value given in (\ref{eq:rstar2}), we have obtained the following estimates
in the even case

\begin{eqnarray}
\nu &=&0.649\,561\,773\,86\,,  \label{eq:WPvpnu} \\
\omega _{1} &=&0.655\,745\,92\,,  \label{eq:WPvpomega1} \\
\omega _{2} &=&3.178\,,  \label{eq:WPvpomega2}
\end{eqnarray}%
where the number of digits has been truncated with regard to the accuracy of
the estimates obtained [by comparison with (\ref{eq:nubest}) and table \ref%
{Table 3}]. We see that the accuracy decreases as the order of the
eigenvalue grows but also that we obtain an estimate of $\omega _{2}$
whereas for that value $\mathrm{v}$ does not vanish at infinity.

The same kind of observations stands in the odd case. We take the
opportunity to indicate that choosing the ADE pair $\left( f,f^{\prime
}\right) $ with $f=$ $\frac{\left( 1-2\lambda \right) }{10}\mathrm{v}$-$x%
\mathrm{v}^{\prime }$ makes \ $f$ vanish for $\lambda >-3/2$ and the
procedure gives a better accuracy on $\breve{\omega}_{1}$ than with the pair 
$\left( \mathrm{v},\mathrm{v}^{\prime }\right) $. This way we obtain the
following estimate [at order $M=20,$ compare with (\ref{eq:omegac})]%
\begin{equation*}
\breve{\omega}_{1}=1.886\,718\,.
\end{equation*}

We have also noted the presence of accumulations of real roots about
spurious positive values of order 5.8 in the even case and 3.77 in the odd
case.

\subsubsection{Litim's eigenvalues}

The determination using the ADE method of the eigenvalues from the Litim
flow equation follows the same lines as previously for the Wilson-Polchinski
flow equation. We limit ourselves in this section to a brief presentation of
the main differences encountered.

\paragraph{ Small field expansion and leading high field behaviour}

Compared to (\ref{eq:VPeqLitim}), we perform a change of eigenfunction, $%
h\rightarrow \,\mathrm{v}_{L}$, according to the symmetry considered:

\begin{itemize}
\item in the even case:%
\begin{equation*}
h\left( \varphi \right) =\,\mathrm{v}_{L}\left( \varphi ^{2}\right) \,,
\end{equation*}

then eq. (\ref{eq:hasy}) yields the following behaviour at large $\bar{x}%
=\varphi ^{2}$:%
\begin{equation*}
\mathrm{v}_{L\text{asy}}\left( \bar{x}\right) =S_{1}{\bar{x}}^{\,\left(
3-\lambda \right) }\left[ 1+O\left( \bar{x}^{-5}\right) \right] \,.
\end{equation*}

\item in the odd case:%
\begin{equation*}
h\left( \varphi \right) =\varphi \,\mathrm{v}_{L}\left( \varphi ^{2}\right)
\,,
\end{equation*}%
and eq. (\ref{eq:hasy}) gives:%
\begin{equation*}
\mathrm{v}_{L\text{asy}}\left( \bar{x}\right) =S_{1}{\bar{x}}^{\,\left(
5/2-\lambda \right) }\left[ 1+O\left( \bar{x}^{-5}\right) \right] \,.
\end{equation*}
\end{itemize}

So defined, the two functions $\mathrm{v}_{L}\left( \bar{x}\right) $ and $%
\mathrm{v}_{L}^{\prime }\left( \bar{x}\right) $ vanish at infinity provided
that $\lambda >3$ in the even case and $\lambda >5/2$ in the odd case
(whereas the arbitrary global normalisation of the eigenfunctions allows to
choose $\mathrm{v}_{L}\left( 0\right) =1$ (even) and $\mathrm{v}_{L}^{\prime
}\left( 0\right) =1$ (odd) corresponding respectively to specific values of $%
S_{1}$).

Although it works, the original ADE pair $\left( \mathrm{v}_{L},\mathrm{v}%
_{L}^{\prime }\right) $ is not the most efficient choice to obtain estimates
of the first nontrivial eigenvalues. A better choice appears to be the pairs 
$\left( f\left( \bar{x}\right) ,f^{\prime }\left( \bar{x}\right) \right) $
with $f\left( \bar{x}\right) =\left( 3-\lambda \right) \mathrm{v}_{L}\left( 
\bar{x}\right) -\bar{x}\mathrm{v}_{L}^{\prime }\left( \bar{x}\right) $ in
the even case and $f\left( \bar{x}\right) =$ $\left( 5/2-\lambda \right) 
\mathrm{v}_{L}\left( \bar{x}\right) -\bar{x}\mathrm{v}_{L}^{\prime }\left( 
\bar{x}\right) $ in the odd case (they correspond to eigenfunctions which
vanish as $\bar{x}\rightarrow \infty $ for more negative values of $\lambda $%
). With these choices and $\bar{k}^{\ast }=0.409\,532\,734\,145\,7$ we
identify immediately the trivial eigenvalues $\lambda _{0}=3$ in the even
case and $\breve{\lambda}_{1}=2.5$, $\breve{\lambda}_{2}=0.5$ in the odd
case but also, for $M=20$, we obtain good estimates of the nontrivial
leading and first subleading eigenvalues:%
\begin{equation*}
\begin{array}{llll}
\nu =\allowbreak 0.649\,561\,774,\quad & \omega _{1}=0.655\,745\,5,\quad & 
\omega _{2}=3.180\,008,\quad & \omega _{3}=5.896, \\ 
\breve{\omega}_{1}=1.886\,703\,7, & \breve{\omega}_{2}=4.524\,1\,, &  & 
\end{array}%
\end{equation*}%
where the numbers of digits have been limited with respect to the estimated
accuracy [compare with (\ref{eq:nubest}), table \ref{Table 3} (even) and (%
\ref{eq:omegac}), table \ref{Table 4} (odd)]. For each eigenvalue, the
successive estimates may be followed unambiguously step by step when $M$
grows so that the right values may be easily selected following the rules
defined\ precedently.

We notice also the presence of spurious convergences and especially in the
even case to the value about 5.8 already encountered with the
Wilson-Polchinski case.

\section{Approximating by hypergeometric functions (HFA)}

\label{Ratios}

The ADE method is most certainly efficient in many cases but it is
relatively heavy regarding the computing time whereas the current methods,
when they work, are lighter. In addition, none of these methods provides a
global solution to the ODE studied: they yield an approximate value of the
integration constant but not a function as global approximation of the
solution looked for.

We propose in this section an alternative method which is lighter than the
ADE method and which provides a global approximation of the solution of
interest. This new method is based on the definition property of the
generalized hypergeometric functions. Let us first review the definition and
main properties of these functions.

\subsection{Generalized hypergeometric functions\label{Brief}}

For $x\in 
\mathbb{C}
$, a series $S=\sum_{n=0}^{\infty }a_{n}x^{n}$ is hypergeometric (see for
example \cite{6181}) if the ratio $a_{n+1}/a_{n}$ is a rational function of $%
n$, i.e.%
\begin{equation*}
\frac{a_{n+1}}{a_{n}}=\frac{P\left( n\right) }{Q\left( n\right) }\,,
\end{equation*}%
for some polynomials $P\left( n\right) $ and $Q\left( n\right) $.

If we factorize the polynomials, we can write:%
\begin{equation}
\frac{a_{n+1}}{a_{n}}=\alpha _{0}\frac{\left( n+\alpha _{1}\right) \left(
n+\alpha _{2}\right) \cdots \left( n+\alpha _{p}\right) }{\left( n+\beta
_{1}\right) \left( n+\beta _{2}\right) \cdots \left( n+\beta _{q}\right)
\left( n+1\right) }\,.  \label{eq:PolyFac}
\end{equation}%
The factor $\left( n+1\right) $ in the denominator may or may not result
from the factorization. If not, we add it along with the compensating factor
in the numerator. Usually, the global factor $\alpha _{0}$ is set equal to 1.

If the set $\left\{ \alpha _{i}\right\} $ includes negative integers, then $%
S $ degenerates into a polynomial in $x.$

When it is not a polynomial, the series $S$ converges absolutely for all $x$
if $p\leq q$ and for $\left\vert x\right\vert <1/\left\vert \alpha
_{0}\right\vert $ if $p=q+1$. It diverges for all $x\neq 0$ if $p>q+1.$

The analytic continuation of the hypergeometric series $S$ with a non-zero
radius of convergence is called a generalized hypergeometric function and is
noted:%
\begin{equation*}
_{p}F_{q}\left( \alpha _{1},\cdots ,\alpha _{p};\beta _{1},\cdots ,\beta
_{q};\alpha _{0}x\right) =\frac{1}{a_{0}}S\,.
\end{equation*}

$_{p}F_{q}\left( x\right) $ is a solution of the following differential
equation (for $\alpha _{0}=1$):%
\begin{equation}
\left[ \theta \left( \theta +\beta _{1}-1\right) \cdots \left( \theta +\beta
_{q}-1\right) -x\left( \theta +\alpha _{1}\right) \cdots \left( \theta
+\alpha _{p}\right) \right] \,_{p}F_{q}\left( x\right) =0\,,
\label{eq:EDOhyper}
\end{equation}%
where%
\begin{equation*}
\theta =x\frac{d}{dx}.
\end{equation*}

When $p>2$ or $q>1$, the differential equation (\ref{eq:EDOhyper}) is of
order $\max \left( p,q+1\right) >2$. It is of second order when $q=1$ and $%
p=0$, $1$ or $2$. It is of first order when $q=0$ and $p=1$

$_{2}F_{1}$ is currently named the hypergeometric function. A number of
generalized hypergeometric functions have also special names: $_{0}F_{1}$ is
called confluent hypergeometric limit function and $_{1}F_{1}$ confluent
hypergeometric function.

In the cases $p\leq q$ for fixed $\left\{ \alpha _{i}\right\} $ and $\left\{
\beta _{i}\right\} $, $_{p}F_{q}\left( x\right) $ is an entire function of $%
x $ and has only one (essential) singular point at $x=\infty $.

For $p=q+1$ and fixed $\left\{ \alpha _{i}\right\} $ and $\left\{ \beta
_{i}\right\} $ in non-polynomial cases\textbf{, }$_{p}F_{q}\left( x\right) $
does not have pole nor essential singularity. It is a single-valued function
on the $x$-plane cut along the interval $\left[ 1,\infty \right] $, i.e. it
has two branch points at $x=1$ and at $x=\infty $.

Considered as a function of $\left\{ \beta _{i};i=1,\cdots ,q\right\} $, $%
_{p}F_{q}\left( x\right) $ has an infinite set of singular points:

\begin{enumerate}
\item $\beta _{i}=-m$, $m\in 
\mathbb{N}
$ which are simple poles

\item $\beta _{i}=\infty $\ which is an essential singular point (the point
of accumulation of the poles).
\end{enumerate}

As a function of $\left\{ \alpha _{i};i=1,\cdots ,p\right\} $, $%
_{p}F_{q}\left( x\right) $ has one essential singularity at each $\alpha
_{i}=\infty $.

The elementary functions and several other important functions in
mathematics and physics are expressible in terms of hypergeometric functions
(for more detail see \cite{6181}).

The wide spread of this family of functions suggests trying to represent the
solution of the ODEs presently of interest in this article, under the form
of a generalized hypergeometric function.

\subsection{\textbf{\ }The HFA method\label{HFA}}

For the sake of the introduction of the new method, let us first consider
the Wilson-Polchinski fixed point equation (\ref{eq:FPw}) and the truncated
expansion (\ref{eq:wmrx}) in which the coefficients $a_{n}\left( r\right) $ $%
(n=1,\cdots ,M)$ are already determined as function of $r$ via a generic
solution of (\ref{eq:FPw}) truncated at order $M$ (in powers of $x)$. The
question is again to construct an auxiliary condition to be imposed\ with a
view to determining the fixed point value $r^{\ast }$. To this end, by
analogy with the generalized hypergeometric property definition recalled in
section \ref{Brief}, we construct the ratio of two polynomials in $n$:%
\begin{equation}
\frac{P_{m_{1}}\left( n\right) }{Q_{m_{2}}\left( n\right) }=\frac{%
\sum_{i=1}^{m_{1}}c_{i}\,n^{i-1}}{\sum_{i=1}^{m_{2}}d_{i}\,n^{i-1}}\,,
\label{eq:hyper0}
\end{equation}%
so that $P_{m_{1}}\left( n\right) /Q_{m_{2}}\left( n\right) $ \ match the $%
M-2$ ratios $a_{n+1}\left( r\right) /a_{n}\left( r\right) $ \ for $%
n=1,\cdots ,M-2$. Hence, accounting for the arbitrariness of the global
normalisation of (\ref{eq:hyper0}), the complete determination of the two
sets of coefficients $\left\{ c_{i};i=1,\cdots ,m_{1}\right\} $ and $\left\{
d_{i};i=1,\cdots ,m_{2}\right\} $ as functions of $r$ implies $%
m_{1}+m_{2}=M-1$.\ Finally, the auxiliary condition on $r$ is obtained by
requiring that the last (still unused) ratio $a_{M}\left( r\right)
/a_{M-1}\left( r\right) $ satisfies again the $n$-dependency satisfied by
its predecessors, namely that:%
\begin{equation}
\frac{\sum_{i=1}^{m_{1}}c_{i}\left( r\right) \left( M-1\right) ^{i-1}}{%
\sum_{i=1}^{m_{2}}d_{i}\left( r\right) \left( M-1\right) ^{i-1}}=\frac{%
a_{M}\left( r\right) }{a_{M-1}\left( r\right) }\,.  \label{eq:hyperauxil}
\end{equation}

The auxiliary condition so obtained is a polynomial in $r$, the roots of
which are candidates to give an estimate at order $M$ of $r^{\ast }$ (noted
below $r_{M}^{\ast }$). Notice that, to obtain faster this auxiliary
condition, one may avoid the calculation of the coefficients $c_{i}\left(
r\right) $\ and $d_{i}\left( r\right) $\ by following the same
considerations as those leading to (\ref{eq:detF}) with the ADE method.

At this point, the method potentially reaches the same goal as the ADE and
other preceding methods. However, according to section \ref{Brief}, in
determining the ratio of polynomials (\ref{eq:hyper0}) we have also
explicitly constructed the function 
\begin{equation}
F_{M}\left( x\right) =r_{M}^{\ast }\cdot \,_{m_{1}+1}F_{m_{2}}\left( \alpha
_{1},\cdots ,\alpha _{m_{1}},1;\beta _{1},\cdots ,\beta _{m_{2}};\alpha
_{0}x\right) \,,  \label{eq:FM}
\end{equation}%
in which $r_{M}^{\ast }$ is the selected estimate of $r^{\ast }$, the sets $%
\left\{ -\alpha _{i}\right\} $ and $\left\{ -\beta _{i}\right\} $ are the
roots of the two polynomials $P_{m_{1}}\left( n\right) $ and $%
Q_{m_{2}}\left( n\right) $ when $r=r_{M}^{\ast }$ whereas: 
\begin{equation}
\alpha _{0}=\frac{c_{m_{1}}\left( r_{M}^{\ast }\right) }{d_{m_{2}}\left(
r_{M}^{\ast }\right) }\,.  \label{eq:alpha0}
\end{equation}

Now, by construction, $F_{M}\left( x\right) $, has the same truncated series
in $x$ as the solution of (\ref{eq:FPw}) we are looking for. This function
is thus a candidate for an approximate representation of this solution.

It is worth noticing that, contrary to the ADE method, the HFA method does
not make an explicit use of the conditions at infinity (large $x$) to
determine $r^{\ast }$. Only a local information, in the neighbourhood of the
origin $x=0$, is explicitly employed.

Let us apply the method to the two equations of interest in this paper.

\subsection{Wilson-Polchinski's equation}

\subsubsection{Fixed point}

We know that the absolute value of the ratio $a_{n}\left( r^{\ast }\right)
/a_{n+1}\left( r^{\ast }\right) $ has a definite value $R_{WP}$ [given by
eq. (\ref{eq:RWP})] as $n\rightarrow \infty $. Consequently, we must
consider the ratio (\ref{eq:hyper0}) with $m_{1}=m_{2}$ (this implies also
that $M$ be odd). In this circumstance, according to section \ref{Brief},
the relevant hypergeometric functions have a branch cut on the positive real
axis (as functions of $\alpha _{0}x$). Consequently the analytic
continuation to large positive values of $x$ is only possible if $\alpha
_{0}<0$. We note also that, according to (\ref{eq:RWP}), $\left\vert \alpha
_{0}\right\vert $ should converge to $1/R_{WP}=0.174774$. Finally by
considering the large $x$ behaviour directly on (\ref{eq:EDOhyper}), it is
easy to convince oneself that the leading power is given by one of the
parameters $\left\{ -\alpha _{i}\right\} $, consequently we expect to
observe a stable convergent value among the $\alpha _{i}$'s toward the
opposite of the leading power at large $x$ of the solution looked for. For
this reason, instead of the function $w\left( x\right) $ of section \ref%
{WPsmallField} the limit of which is 1 as $x\rightarrow \infty $ [see (\ref%
{eq:asyw})], we have considered the translated function $w_{t}\left(
x\right) =w\left( x\right) -1$ which, according to eqs (\ref{eq:fasy}) and (%
\ref{eq:chgt1}, \ref{eq:astar}), tends to $A^{\ast }x^{-2/5}$. In this case
we\ thus expect to observe a stable value among the $\alpha _{i}$'s about $%
0.4$ with the eventual possibility of estimating $A^{\ast }$.

\begin{figure}[tbp]
\begin{center}
\includegraphics*[width=10cm]{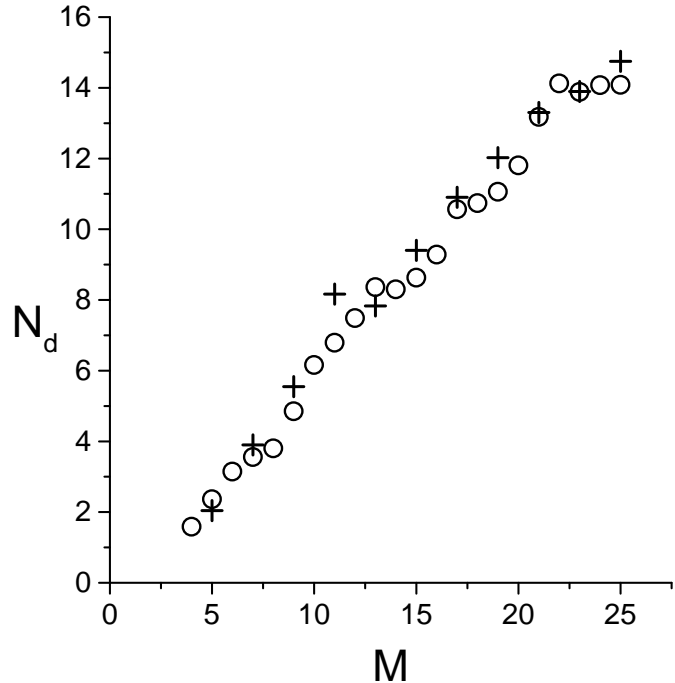}
\end{center}
\caption{Respective approximate number of digits (defined in the caption of
figure \protect\ref{fig5}) obtained for r$_{M}^{\ast }$ with the HFA method
(crosses) and the ADE method (open circles) for the Wilson-Polchinski fixed
point equation. Whereas at a given order $M$ the accuracy is similar, a
smaller time computing is necessary with the HFA method.}
\label{fig6}
\end{figure}

When looking at the roots of the auxiliary condition (\ref{eq:hyperauxil})
as $M$ varies, we obtain the same kind of accumulations about the expected
fixed point value $r^{\ast }$ as shown in figure \ref{fig2} (with much less
points however). We can also easily select the right nontrivial solution
using the procedure described just above (\ref{eq:rstar2}). We get precisely
this excellent estimate with $M=25$ and a reduced computing time compare to
the ADE method. Figure (\ref{fig6}) shows the accuracies obtained on $%
r^{\ast }$ (crosses) compared to the ADE method (open circles). 

Furthermore, the sets of parameters of the successive hypergeometric
functions involve two stable quantities the values of which at $M=25$ are:%
\begin{eqnarray}
\alpha _{0} &=&-0.174\,775\,,  \label{eq:alpha0estim} \\
\alpha _{1} &=&0.396\,2\,.  \label{eq:alpha1estim}
\end{eqnarray}

Those two results are quantitatively and qualitatively very close to the
expected values (respectively $-0.174774$ and $0.4$ as given just above).

This clearly shows that the hypergeometric function determined this way
provides us with a really correct (but approximate) global representation of
the fixed point function. This contrasts strongly with the numerical
integration of the ODE which, due to the presence of the moving singularity,
never provides us with such an approximate global representation of the
solution looked for.

From (\ref{eq:alpha1estim}) we have obtained a rough estimate of $A^{\ast }$
($=6b^{\ast }/5$) by a direct consideration of the value of the
corresponding function\ $F_{M}\left( x\right) $ defined in (\ref{eq:FM}) for
some relatively large value of $x$ and we obtain $A^{\ast }\simeq -2.6$ what
is a sufficiently accurate estimate to serve as a guess in the shooting
method.

We have also tried to determine, using the HFA method, the value $A^{\ast }$
directly from the \textquotedblleft reverse side\textquotedblright\
corresponding to (\ref{eq:WFPasy}). We have not improved the previous\
biased estimate obtained by ADE (about $A^{\ast }=-2.735$). We do not
understand the significance of this coincidence. We recall, however, that
the radius of convergence of the Taylor series of $u^{\ast }\left( y\right) $
about $y=0$ probably vanishes. This biased result shows again that the
property of convergence of the Taylor series is crucial for the accuracy of
the two methods.

\subsubsection{Eigenvalues}

We have also applied the HFA method to the determination of the eigenvalues.
With $M=17$, we have easily and without ambiguity obtained the following
excellent estimates [compare with (\ref{eq:nubest}, \ref{eq:omegac}) and
tables \ref{Table 3} and \ref{Table 4}]:%
\begin{equation*}
\begin{array}{lll}
\nu =\allowbreak \allowbreak 0.649\,561\,774\,,\quad & \omega
_{1}=0.655\,745\,939\,3\,,\quad & \omega _{2}=3.180\,006\,53\,, \\ 
\omega _{3}=5.912\,229\,4\,,\quad & \omega _{4}=8.796\,045\,,\quad & \omega
_{5}=11.800\,4\,, \\ 
\breve{\omega}_{1}=1.886\,703\,839\,,\quad & \breve{\omega}%
_{2}=4.524\,390\,3\,,\quad & \breve{\omega}_{3}=7.337\,635\,.%
\end{array}%
\end{equation*}

These results show a greater efficiency than with the ADE method especially
in the determination of the subleading eigenvalues.

It is worth indicating also that, surprisingly enough, we observe again
(i.e. as with the ADE method) the presence of convergences to the same
spurious eigenvalues: 5.8 and 3.8 in the even and odd cases respectively.

\subsection{Litim's equation\label{GeoLitim}}

\subsubsection{Fixed point}

Applying the HFA method with the ratio of two successive coefficients $%
a_{n}\left( \bar{k}\right) $ provides again an accumulation of roots about
the right value of $\bar{k}^{\ast }$ given in (\ref{eq:kstar}). However,
this time, we have encountered some difficulties in defining a process of
selection of the right root. We obtain the following estimate for $M=21$:%
\begin{equation*}
\bar{k}^{\ast }\simeq 0.409\,531\,,
\end{equation*}%
which is not bad [compare with (\ref{eq:kstar})] but not as satisfactory as
in the preceding Wilson-Polchinski's case.

With regard to the transformation (\ref{eq:Legendre}) and the preceding
success of the HFA method, it is not amazing that the representation of the
solution in the Litim case be more complicated than in the Wilson-Polchinski
case.

We have already mentioned that, instead of the ratio of two successive terms
of the series $a_{n}\left( \bar{k}\right) $, it is a shifted ratio that
roughly converges to the finite radius of convergence (\ref{eq:radiusL}).\
As a matter of fact, if we use the ratios%
\begin{equation*}
\frac{a_{n+3}\left( \bar{k}\right) }{a_{n}\left( \bar{k}\right) }\,,
\end{equation*}%
instead of the ratio $a_{n+1}/a_{n}$ without changing the procedure\footnote{%
Notice that the procedure does not define some generalized hypergeometric
function of $\bar{x}^{3}.$ This would have been obtained by considering
separately three series in the original series. Then a combination of three
generalized hypergeometric functions would have represented the solution
looked for.} described in section \ref{HFA}, then we get a better estimate
for $M=21$ [compare with (\ref{eq:kstar})]:%
\begin{equation*}
\bar{k}^{\ast }\simeq 0.409\,532\,737\,,
\end{equation*}%
although the convergence properties are not substantially modified.

Because the case is apparently more complicated than precedently, we do not
pursued further the discussion of the global representation of the fixed
point solution by generalized hypergeometric functions.

\subsubsection{Eigenvalues}

For the eigenvalue problem, a similar difficulty occurs where the right
values do not appear as clear convergent series of roots. At order $M=17$,
we get the following estimates:$\allowbreak $%
\begin{equation*}
\begin{array}{llll}
\nu =0.649\,55\,,\quad & \omega _{1}=0.657\,6\,,\quad & \omega
_{2}=3.20\,,\quad & \omega _{3}=5.8\,, \\ 
\breve{\omega}_{1}=1.89\,,\quad & \breve{\omega}_{2}=4.5\,. &  & 
\end{array}%
\end{equation*}

As in the case of the fixed point determination, if instead of applying the
method with the ratio of two successive terms of the series $a_{n}\left( 
\bar{k}\right) $ we consider the ratios%
\begin{equation*}
\frac{a_{n+3}\left( \bar{k}\right) }{a_{n}\left( \bar{k}\right) }\,,
\end{equation*}%
then we get better estimates for $M=19$:%
\begin{equation*}
\begin{array}{llll}
\nu =0.649\,561\,774\,,\quad & \omega _{1}=0.655\,75\,,\quad & \omega
_{2}=3.180\,7\,,\quad & \omega _{3}=5.905\,, \\ 
\breve{\omega}_{1}=1.886\,71\,,\quad & \breve{\omega}_{2}=4.524\,. &  & 
\end{array}%
\end{equation*}%
where the numbers of digits have been limited having regard to the estimated
accuracies [compare with (\ref{eq:nubest}), table \ref{Table 3} (even) and (%
\ref{eq:omegac}), table \ref{Table 4} (odd)].

\section{Summary and conclusions}

\label{Conc}

We have presented the details of a highly accurate determination of the
fixed point and the eigenvalues for two equivalent ERGEs in the local
potential approximation. First, we have made use of a standard numerical
(shooting) method to integrate the ODEs concerned. Beyond the test of the
equivalence between the two equations, already published in \cite{6137}, the
resulting numerics have been used to concretely test the efficiency of two
new approximate analytic methods for solving two point boundary value
problems of ODEs based on the expansion about the origin of the solution
looked for (field expansion).

We have considered explicitly those two methods applied to the study of the
two equivalent ODEs. We have shown that they yield estimates as accurate as
those obtained with the shooting method provided that the Taylor series
about the origin of the function looked for has a non-zero radius of
convergence.

This is an important new result since, up to now, no such approximate
analytical method was known to work in the simplest case of the
Wilson-Polchinski equation. In the case of the Litim equation the two
methods converge better than the currently used expansions (usually referred
to as I and II in the literature, see e.g. [\cite{3553}]). Our results
support concretely the conclusions of \cite{5902} which indicated that the
high field contributions were important in the Wilson-Polchinski case
whereas they were less important in the Litim case.

The first of the two methods relies upon the construction of an auxiliary
differential equation (ADE) satisfied by the Taylor series at the origin and
to which is imposed the condition of the second boundary (at infinity) \cite%
{6110}.

The second method (HFA) is new. It consists in defining a global
representation of the solution of the ODE via a generalized hypergeometric
function. The HFA method provides the advantage of yielding a global
(approximate) representation of the solution via an explicit hypergeometric
function.

In both cases it is possible to obtain easily (with few terms in the field
expansion) rough estimates of the solution which may be used as guesses in a
subsequent shooting method.

The procedures may be applied to several coupled ODEs as shown in \cite{6110}
for the ADE method. Hence, we hope that the present work will\ make easier
and more efficient future explicit (and ambitious) considerations of the
derivative expansion of exact renormalisation group equations.

\section{Acknowledgements}

We thank D. Litim for comments on an earlier version of this article.


\begin{thebibliography}{99}
\bibitem{206} M. Gell-Mann and F. E. Low, Phys. Rev. \textbf{95} (1954) 1300.

K. G. Wilson, Phys. Rev. \textbf{140} (1965) B445; \emph{ibid.} D \textbf{2}
(1970) 1438.

\bibitem{425} K. G. Wilson, Rev. Mod. Phys. \textbf{47} (1975) 773.

\bibitem{4948} J. Zinn-Justin, Euclidean Field Theory and Critical
Phenomena, Fourth edition\ (Clarendon Press, Oxford, 2002).

A. Pelissetto and E. Vicari, Phys. Rep. \textbf{368} (2002) 549.

\bibitem{4374} C. Wetterich, Phys. Lett. B \textbf{\ 301} (1993) 90.

M. Bonini, M. D'Attanasio and G. Marchesini, Nucl. Phys. B \textbf{409}
(1993) 441.

T. R. Morris, Int. J. Mod. Phys. A\textbf{\ 9} (1994) 2411.

U. Ellwanger, Z. Phys. C\textbf{\ 62} (1994) 503.

\bibitem{Irvine} K.G. Wilson, Irvine Conference, 1970, unpublished. See the
equation in K.~G.~Wilson and J.~B.~Kogut, Phys.\ Rep.\ \textbf{12} (1974) 75.

\bibitem{414} F. J. Wegner and A. Houghton, Phys. Rev. A\textbf{\ 8} (1973)
401.

\bibitem{354} J. Polchinski, Nucl. Phys. B \textbf{231} (1984) 269.

\bibitem{4595} K. I. Aoki, Int. J. Mod. Phys. B\textbf{\ 14} (2000) 1249.

C. Bagnuls and C. Bervillier, Phys. Rep. \textbf{348} (2001) 91.

J. Berges, N. Tetradis and C. Wetterich, Phys. Rep. \textbf{363} (2002) 223.

J. Polonyi, Cent. Eur. J. Phys. \textbf{1} (2003) 1.

For a recent overview of advanced functional RG methods, see
J.~M.~Pawlowski, hep-th/0512261.

For a pedagogical introduction see B. Delamotte, cond-mat/0702365.

\bibitem{4281} J. F. Nicoll and T. S. Chang, Phys. Lett. A \textbf{62}
(1977) 287.

\bibitem{212} G. R. Golner, Phys. Rev. B\textbf{\ 33} (1986) 7863.

\bibitem{3478} A. Margaritis, G. \'{O}dor and A. Patk\'{o}s, Z. Phys. C%
\textbf{\ 39} (1988) 109. Actually the procedure followed there had already
been implemented in another context by F. M. Fernandez and E. A. Castro, J.
Phys. A 14 (1981) L485 and by J. R. Silva and S. Canuto, Phys. Lett. A 88
(1982) 282; ibid. A 101 (1984) 326; ibid. Phys. A 106 (1984) 1.

\bibitem{3642} N. Tetradis and C. Wetterich, Nucl. Phys. B\textbf{\ 422}
(1994) 541.

\bibitem{4192} M. Alford, Phys. Lett. B\textbf{\ 336} (1994) 237.

\bibitem{3553} K. I. Aoki, K. Morikawa, W. Souma, J. I. Sumi and H. Terao,
Prog. Theor. Phys. \textbf{95} (1996) 409; \emph{ibid.} \textbf{99} (1998)
451.

\bibitem{6110} B. Boisseau, P. Forgacs and H. Giacomini, J. Phys. A \textbf{%
40} (2007) F215. 

\bibitem{6201} P. Amore and F. M. Fernandez, arXiv:0705.3862 (2007).

\bibitem{5020} D. F. Litim, Phys. Rev. D \textbf{64}, 105007 (2001); Phys.
Lett. B \textbf{486} (2000) 92.

\bibitem{5049} D. F. Litim, Int. J. Mod. Phys. A \textbf{16} (2001) 2081.

\bibitem{5902} D. F. Litim, J. High Energy Phys. \textbf{07} (2005) 005.

\bibitem{5911} T. R. Morris, J. High Energy Phys. \textbf{07} (2005) 027.

\bibitem{6137} C. Bervillier, A. J\"{u}ttner and D. F. Litim, Nucl. Phys. B 
\textbf{783} (2007) 213.

\bibitem{2080} G. Felder, Comm. Math. Phys. \textbf{111} (1987) 101.

\bibitem{439} K. G. Wilson and M. E. Fisher, Phys. Rev. Lett. \textbf{28}
(1972) 240.

\bibitem{5465} W. H. Press, S. A. Teukolsky, W. T. Vetterling and B. P.
Flannery, The Art of Scientific Computing\ (Cambridge University Press,
1992).

\bibitem{3491} R. D. Ball, P. E. Haagensen, J. I. Latorre and E. Moreno,
Phys. Lett. B\textbf{\ 347} (1995) 80.

J. Comellas and A. Travesset, Nucl. Phys. B\textbf{\ 498} (1997) 539.

C. Bervillier, Phys. Lett. A \textbf{332} (2004) 93.

\bibitem{5252} D. F. Litim, Nucl. Phys. B\textbf{\ 631} (2002) 128.

\bibitem{5625} D. F. Litim and L. Vergara, Phys. Lett. B\textbf{\ 581}
(2004) 263.

\bibitem{5469} L. Canet, B. Delamotte, D. Mouhanna and J. Vidal, Phys. Rev.
B \textbf{68} (2003) 064421; \emph{ibid.} D\textbf{\ 67} (2003) 065004.

\bibitem{5903} J. P. Blaizot, R. Mendez-Galain and N. Wschebor, Phys. Lett.
B \textbf{632} (2006) 571; see also hep-th/0605252.

D. Guerra, R. Mendez-Galain and N. Wschebor, arXiv:0704.0258.

\bibitem{5677} B. Delamotte, D. Mouhanna and M. Tissier, J. Phys.: Condens.
Matter \textbf{16} (2004) S883; Phys. Rev. B\textbf{\ 69} (2004) 134413.

\bibitem{3358} T. R. Morris, Phys. Lett. B \textbf{334} (1994) 355.

\bibitem{6181} G. E. Andrews, R. Askey and R. Roy, Special Functions\
(Cambridge University Press, 1999).

G. Gasper and M. Rahman, Basic Hypergeometric Series\ (Cambridge University
Press, 2004).
\end{thebibliography}
\end{document}